\documentclass[aps,pra,twocolumn,amsmath,amssymb,nofootinbib,showpacs,longbibliography,superscriptaddress]{revtex4-2}

\usepackage{amsmath}
\usepackage{amssymb}
\usepackage{graphicx}
\usepackage{color}
\usepackage{float}
\usepackage[english]{babel} 

\usepackage[usenames,dvipsnames]{xcolor}
\usepackage[normalem]{ulem} % provides \sout
\usepackage[colorlinks=true,citecolor=Cerulean,linkcolor=RubineRed,urlcolor=Cerulean]{hyperref}

\newcommand{\rr}{\mathbf{r}}
\newcommand{\asc}{a_{\textrm{sc}}}
\newcommand{\phig}{\phi_{\textrm{g}}}
\newcommand{\tu}{\theta_1}
\newcommand{\td}{\theta_2}

\newcommand{\dng}[1]{#1}
\newcommand{\correct}[1]{}

\newcommand{\kB}{k_{\textrm{\tiny B}}}
\newcommand{\Er}{E_{\textrm{r}}}
\newcommand{\fs}{f_{\textrm{s}}}

\newcommand{\HBHm}{\hat{H}_{\textrm{\tiny BH}}}
\newcommand{\JA}{J'_{\textrm{A}}}
\newcommand{\JB}{J'_{\textrm{B}}}
\newcommand{\JAB}{J'_{\textrm{A/B}}}
\newcommand{\DeltaAB}{\Delta_{\textrm{AB}}}
\newcommand{\UB}{U_{\textrm{B}}}
\newcommand{\UA}{U_{\textrm{A}}}
\newcommand{\UAB}{U_{\textrm{A/B}}}
\newcommand{\wA}{w_{\textrm{A}}}
\newcommand{\wB}{w_{\textrm{B}}}
\newcommand{\wAB}{w_{\textrm{A/B}}}

\newcommand{\Uc}{U_\textrm{c}}
\newcommand{\Tc}{T_\textrm{c}}
\newcommand{\ns}{n_\textrm{s}}
\newcommand{\lambdaT}{\lambda_T}
\newcommand{\lambdaTc}{\lambda_{\Tc}}

\newcommand{\tgTwoD}{\tilde{g}_0}

\begin{document}

\title{Quantum phase diagrams for bosons in hexagonal optical potentials: A continuous-space quantum Monte Carlo study}

\author{Danilo Nascimento Guimaraes}
\affiliation{CPHT, CNRS, École Polytechnique, Institut Polytechnique de Paris, Palaiseau, France}

\author{Laurent Sanchez-Palencia}
\affiliation{CPHT, CNRS, École Polytechnique, Institut Polytechnique de Paris, Palaiseau, France}

\date{\today}

\begin{abstract}
Hexagonal optical lattices, emulating graphene and hexagonal boron nitride (h-BN) structures, provide a versatile platform for exploring strongly correlated quantum matter. Using continuous-space exact diagonalization and quantum Monte Carlo simulations, we investigate the phase diagrams of ultracold bosons in honeycomb and h-BN lattices. For the honeycomb lattice, we find significant deviations from the standard Bose--Hubbard model even for strong lattice amplitudes. We observe suppressed Mott insulator lobes and the absence of higher-order insulating phases, \correct{attributed to strong density-assisted tunneling effects} \dng{which we attribute to strong density-assisted tunneling effects}. In the h-BN case, a rich phase diagram emerges, featuring multiple Mott lobes with various sublattice occupations, driven by the interplay of lattice asymmetry, interactions, and particle filling. Our results highlight the necessity of continuous-space treatments for capturing the full complexity of bosonic quantum phases in hexagonal geometries, paving the way for experimental realizations with ultracold atoms and further theoretical work.
\end{abstract}

\maketitle

\section{Introduction}
Solid-state materials exhibit a remarkable diversity of geometric structures, arising from the interplay of atomic electronic configurations and complex interactions. Beyond conventional cubic lattices, these include triangular, checkerboard, and honeycomb lattices, as well as quasicrystals, and multilayer systems~\cite{kittel2004,janot1994,Castro2009,bistritzer2011,andrei2021}.
Such structures give rise to a wealth of nontrivial phenomena ranging from exotic superconductivity and anomalous quantum Hall effects to fractal behavior, localization, and topological phases, which are currently attracting considerable attention~\cite{haldane1988,hasanKane2010,kamiya2018,uri2023,rolof2013,evers2008,huang2018}.
The challenge lies in unraveling the intricate interplay of matter-wave interference, electronic interactions, and geometric effects that underpin these phenomena.
In parallel, ultracold atomic systems have emerged as a versatile platform for quantum simulation, offering unprecedented control over interaction strengths, sign, and range, as well as, lattice geometries and dynamical parameters~\cite{lewenstein2007,bloch2008,bloch2012,gross2017,esslinger2010,tarruell2018,*lsp2018,modugno2010,lsp2010}. By engineering laser configurations, exotic lattice structure, such as triangular, honeycomb, Lieb, and Kagome lattices, as well as super-lattices and quasiperiodic lattices, can be emulated~\cite{jaksch1998,windpassinger2013,wirth2011,becker2010,struck2011,tarruell2012,flaschner2016,taie2015,santos2004,jo2012,anderlini2007,trotzky2008,lsp2005,jagannathan2013,viebahn2019}.
These systems allow us to not only revisit the physics of strongly correlated electronic systems but also explore new physics of bosonic quantum matter in such structures.
Moreover, recent advances, including multi-frequency lattices and dynamical phase control, further expand the range of accessible geometries~\cite{kosch2022,brown2022,shimasaki2024,neely2026}.

Hexagonal lattices, exemplified by graphene and certain carbon nanotubes~\cite{novoselov2004,Castro2009}, are of particular interest due to their unique band structures, which feature Dirac cones, flat bands, and exaggerated correlation effects, as well as their rich topological properties~\cite{geim2007,miao2022,hirata2021,Castro2009}. Up to date only a few experiments have realized hexagonal lattices but they have already facilitated studies of exotic effects, including multi-orbital superconductivity~\cite{soltanPanahi2012}, antiferromagnetic N\'eel ordering~\cite{soltanPanahi2011}, and Dirac point physics~\cite{tarruell2012}. It is also worth mentioning cavity polariton systems with nanopillars arrays arranged in honeycomb geometry~\cite{stJean2021,real2020}.
Theoretically, tight-binding models (TBMs) have been extensively employed to describe the single-particle band structure of honeycomb and gapped-honeycomb systems~\cite{Wallace1947,Castro2009,anisimovas2014,kuskBlock2010}, while Bose--Hubbard models \dng{(BHMs)} have been used to study superfluid \dng{(SF)}--Mott insulator \dng{(MI)} transitions and superlattice-induced insulating phases~\cite{Teichmann2010,PhysRevA.85.023619,wang2025}.
Beyond, dynamical control of optical lattices has enabled the realization of hexagonal boron nitride (h-BN, aka gapped graphene) models~\cite{kosch2022}, with recent theoretical work revealing transitions between MI, checkerboard and SF phases at unit filling, tunable by the lattice gap~\cite{wang2025}.

In this work, we investigate the dynamics of ultracold bosons in hexagonal optical potentials, including honeycomb and h-BN lattices. Using continuous-space calculations, we determine the band structure of these models, establishing the validity limits of TBM approximations and extracting their parameters. Moreover, quantum path-integral Monte Carlo (PIMC) simulations yield exact phase diagrams at vanishingly small temperatures and arbitrary fillings. For the honeycomb lattice, we find standard SF and MI phases. However, we observe significant deviations from BHMs, even at strong lattice amplitudes where band structures are very well reproduced. For instance, for a lattice depth $V_0=15\,\Er$, the lowest band is reproduced by the simplest TBM with an accuracy of about 1\% but the BHM approximation fails to describe but the first Mott lobe. \correct{These deviations are attributed to strong density-assisted tunneling effects beyond the BHM framework} \dng{Mean-field estimates indicate that density-assisted tunneling effects are very strong and likely account for the observed deviations}. For the h-BN lattice, our results unveil a rich phase diagram characterized by multiple Mott lobes with various fillings, hence extending the findings of Ref.~\cite{wang2025} and highlighting the interplay of interactions, filling, and lattice geometry.
Our work paves the way for experiments with ultracold atoms in hexagonal optical lattices and motivates further exploration of multilayer systems.

\section{Model} 
\begin{figure}[t!]
    \centering
    \includegraphics[width=0.75\linewidth]{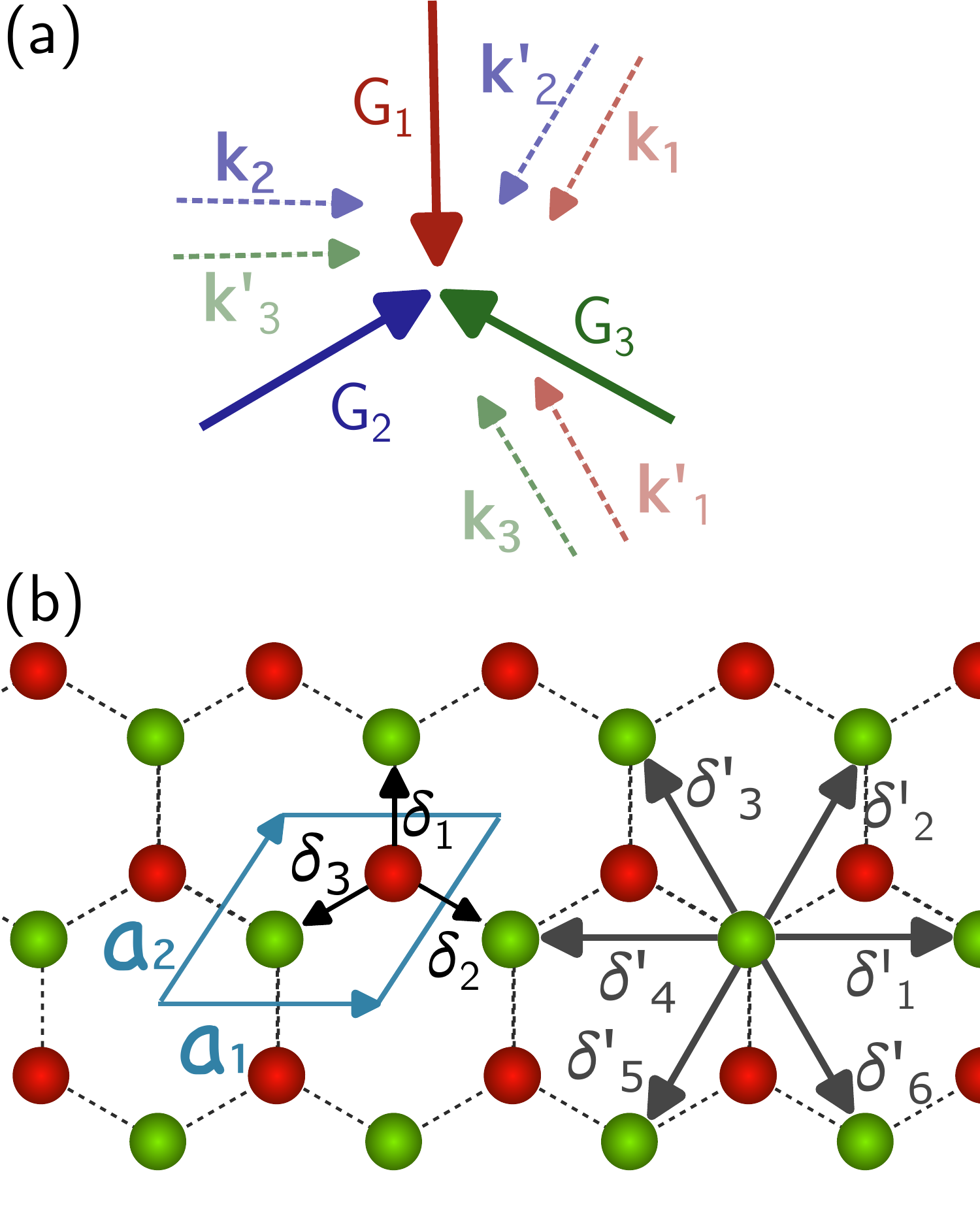}
    \caption{
Hexagonal lattice.
(a)~Lattice vectors $\mathbf{G}_j$ entering the potential of Eq.~\eqref{eq:potential}.
Faint arrows indicate the laser wave vectors in the implementation of Ref.~\cite{kosch2022}. 
(b)~Lattice formed by the minima of the potential in Eq.~\eqref{eq:potential}. It is composed of two interpenetrating triangular sublattices A (red) and B (green). Also shown are the Bravais vectors $\mathbf a_{1,2}$, the nearest-neighbor vectors $\boldsymbol{\delta}_j$, and the next-nearest-neighbor vectors $\boldsymbol{\delta}'_j$.}
    \label{fig:lattices}
\end{figure}

The dynamics of a 2D gas of identical spinless bosons with mass $m$ and repulsive two-body interactions is governed by the Hamiltonian
%-------------------------------%
\begin{equation}\label{eq:Hamiltonian}
\hat{H} 
=
\int d\rr\,
\hat{\Psi}^\dagger \left[\hat{H}_0 + \frac{1}{2} \int d\rr'\, \hat{\Psi}'\,^\dagger U(\rr-\rr') \hat{\Psi}'\right]\hat{\Psi},
\end{equation}
%-------------------------------%
where $\hat{\Psi}$ and $\hat{\Psi}'$ are, respectively, the field operators at positions $\rr$ and $\rr'$, 
$\hat{H}_0=-{\hbar^2\nabla^2}/{2m}+V(\rr)$ is the one-body Hamiltonian with potential $V(\rr)$, and $U(\rr-\rr')$ is the two-body interaction potential.
We consider the honeycomb-like lattice potential
%-------------------------------%
\begin{equation}\label{eq:potential}
    V(\mathbf{r}) = 2V_0 \sum_{j=1}^{3} \cos(\mathbf{G}_j \cdot \mathbf{r} + \phi_j)\,,
\end{equation}
%-------------------------------%
where $V_0$ is the potential amplitude, the $\phi_j$'s are phases, and the $\mathbf{G}_j$'s are vectors with size $\pi/a$ and arranged to form angles of $2\pi/3$ with each other.
A possible implementation of such a potential with controlled parameters involves using multifrequency settings with three pairs of lasers~\cite{kosch2022}, see Fig.~\ref{fig:lattices}(a).
We then have $\mathbf{G}_j=\mathbf{k}_j-\mathbf{k}'_j$,
where $\mathbf{k}_j$ and $\mathbf{k}'_j$ are the laser wavevectors with $|\mathbf{k}_j|=|\mathbf{k}'_j|=k$, and $\phi_j$ is the phase difference for beams in pair $j$. Since $|\mathbf{G}_j|=\sqrt{3}\,k$, the choice $|\mathbf{G}_j|=\pi/a$ implies $k=\pi/(\sqrt{3}a)$. In what follows, we use the $a=\pi/\sqrt{3}k$ as the length unit and
recoil energy
\begin{equation}
    \Er=\frac{\hbar^2 k^2}{2m}=\frac{\pi^2\hbar^2}{6ma^2},
\end{equation}
as the energy unit.
Note that we can restrict ourselves to $V_0 \geq 0$ since changing the sign of $V_0$ is equivalent to shifting each $\phi_j$ into $\phi_j+\pi$. The choice of the spatial origin provides two independent degrees of freedom, allowing two of the three phases to be absorbed by a coordinate shift. As a result, the geometry of the potential is determined by a single phase. Without loss of generality, we may choose the so-called \textit{geometric phase} $\phig=\sum_j\phi_j$~\cite{kosch2022} and take $\phi_1=\phi_2=\phi_3=\phig/3$. \dng{The zero of energy is taken to be the ground-state energy of the corresponding single-particle Hamiltonian $\hat H_0$ for the parameters $(V_0,\phig)$ under consideration.}

Irrespective of the value of $\phig$, the potential $V(\rr)$ is periodic with Bravais vectors $\mathbf{a}_1=(4a/\sqrt{3})\mathbf{e}_x$ and $\mathbf{a}_2=(2a/\sqrt{3})(\mathbf{e}_x+\sqrt{3}\mathbf{e}_y)$, with $\mathbf{e}_x,\mathbf{e}_y$ the canonical basis of $\mathbb{R}^2$, see Fig.~\ref{fig:lattices}(b). 
For $\phig=0$, the potential minima form a honeycomb lattice with identical A and B sites, \dng{represented, respectively, by red and green spots in Fig.~\ref{fig:lattices}(b)}.
An increasing value of $\phig$ progressively lifts the degeneracy between A and B sites, hence forming an hBN-like potential. For $|\phig|>\pi/2$, the B points are no longer local minima while the A sites are still minima, hence forming a triangular lattice, see Appendix~\ref{app:Minima}.

In the low-energy s-wave scattering regime, the interaction potential is characterized by the sole scattering length $a_{\rm sc}$ and, for weak enough interactions, $U(\mathbf r)$ may be replaced by a contact interaction, $U(\rr)=(\hbar^2\tilde{g}/m)\delta(\rr)$,
with dimensionless coupling parameter
\begin{equation}
    \tilde{g}\simeq 
    \frac{1}{
      \tilde{g}_0^{-1}+(4\pi)^{-1}\ln(\Lambda \Er/\mu)
    }, \label{eq:gRenorm}
\end{equation}
with
\begin{equation}
    \tilde{g}_0=\frac{2\pi}{
\ln(a/a_{\rm sc})},
\end{equation}
$\mu$ the chemical potential, and $\Lambda\simeq0.423$ a numerical constant\footnote{With our convention \(\Er=\hbar^2k^2/2m=\pi^2\hbar^2/(6ma^2)\) the recoil-energy unit differs by a factor of \(3\) from the more common convention \(\Er=\pi^2\hbar^2/(2ma^2)\). Accordingly, the constant \(\Lambda\) in Eq.~\eqref{eq:gRenorm} is three times the value given in Refs.~\cite{bloch2008,Petrov2000,Petrov2001,Pricoupenko2007,gautier2021}.}~\cite{bloch2008,Petrov2000,Petrov2001,Pricoupenko2007,gautier2021}.
This relation holds for homogeneous two-dimensional systems in the weakly-interacting regime $\tilde{g} \ll 1$
and is used here heuristically. 

%-------------------------------%
%-------------------------------%

%-------------------------------%
\begin{figure*}[t!]
    \centering
    \includegraphics[width=\linewidth]{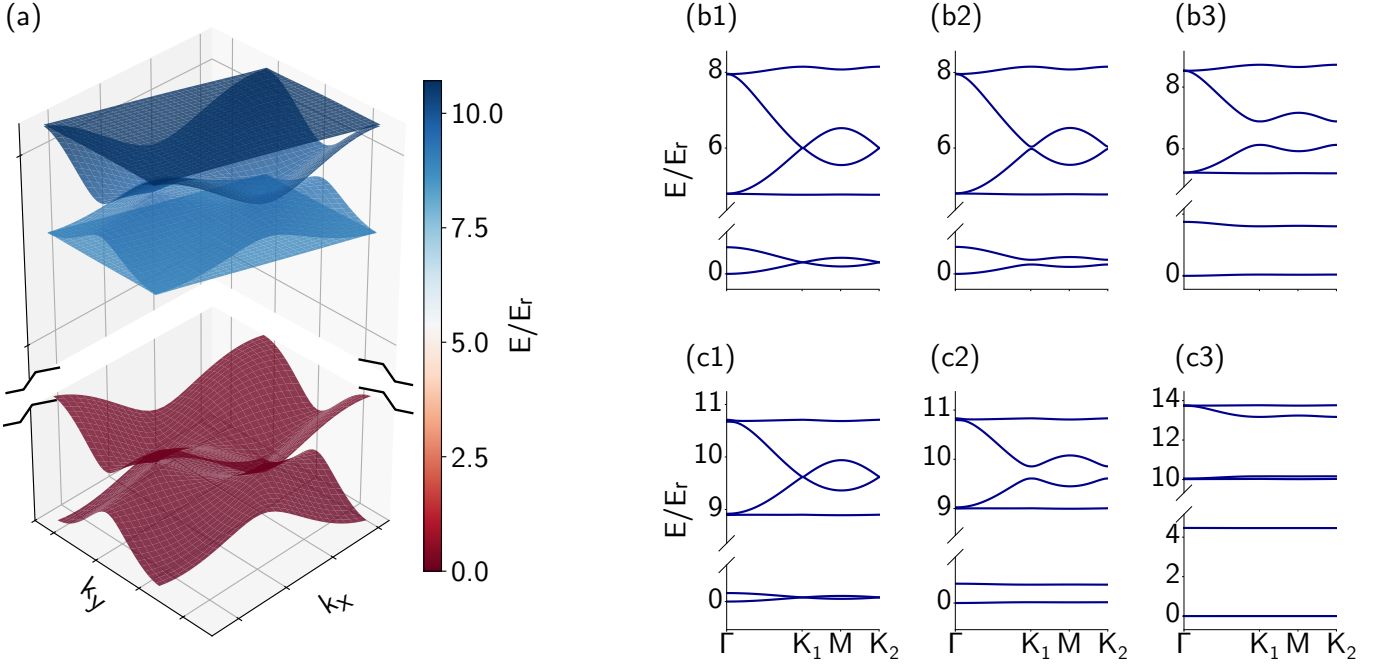}
    \caption{
    Single-particle band structure for a honeycomb potential with various amplitudes and geometric phases.
    Results obtained using exact diagonalization of Hamiltonian $\hat H_0$.
    (a)~Full dispersion relations in the first Brillouin zone for the first bands, $V_0=15\,\Er$, and $\phig=0$.
    (b1) Band dispersion relations along the high-symmetry path $\Gamma$–$K_1$–$M$–$K_2$ for $V_0=5\,\Er$, and
    (b1)~$\phig=0$, (b2)~$\phig=0.008$, and (b3)~$\phig=0.1$.
    (c)~Same as~(b) but for $V_0=15\,\Er$.
    }
    \label{fig:bands}
\end{figure*}
%-------------------------------%

For sufficiently deep lattice potential, $V_0 \gg \Er$, and low-energy, 
the system may be described, up to an irrelevant constant, by the extended BHM Hamiltonian
%-------------------------------%
\begin{eqnarray}
    \HBHm & = &
    -J\sum_{\langle j,\ell\rangle}
    \hat{a}_j^\dagger \hat{a}_\ell
    - \frac{\DeltaAB}{2}\sum_{j\in A}\hat{a}_j^\dagger \hat{a}_j
    + \frac{\DeltaAB}{2}\sum_{j\in B}\hat{a}_j^\dagger \hat{a}_j
    \nonumber\\
    && -\JA\sum_{\langle\langle j,\ell \rangle\rangle \in A} \hat{a}_j^\dagger \hat{a}_\ell
       -\JB\sum_{\langle\langle j,\ell \rangle\rangle \in B} \hat{a}_j^\dagger \hat{a}_\ell
    \label{eq:TBM.Hamiltonian} \\
    && +\frac{\UA}{2}\sum_{j\in A}
    \hat{n}_j\left(\hat{n}_j - 1\right) +\frac{\UB}{2}\sum_{j\in B} \hat{n}_j\left(\hat{n}_j - 1\right),
    \nonumber
\end{eqnarray}
%-------------------------------%
where $\hat{a}_j$, $\hat{a}_j^\dagger$, and $\hat{n}_j=\hat{a}_j^\dagger\hat{a}_j$ are, respectively, the annihilation, creation, and number operators for bosons in site $j$,
$J$ is the tunneling energy between nearest-neighbor (nn) A-B sites,
$\JA$ and $\JB$ are the tunneling energies between next-nearest-neighbor (nnn) A-A or B-B sites,
$\DeltaAB \geq 0$ is the energy difference between A and B sites\footnote{By convention, we choose A sites as having lower onsite energy.},
and $\UA$ and $\UB$ are, respectively, the interaction energy in A and B sites.
The notations $\langle j,\ell \rangle$ and $\langle\langle j,\ell \rangle\rangle$ refer to sums over pairs of
nn A-B sites and pairs of nnn A-A or B-B sites, respectively.

\section{Band structure and tight-binding parameters}
We first study the single-particle part of Eq.~\eqref{eq:Hamiltonian}, associated with the Hamiltonian $\hat{H}_0$.
Using Bloch's theorem, $\hat{H}_0$ may be diagonalized writing the wavefunctions $\psi_{b,\mathbf{k}}(\mathbf{r})=u_{b,\mathbf{k}}(\mathbf{r})e^{i\mathbf{k}\cdot\mathbf{r}}$, where $u_{b,\mathbf{k}}(\mathbf{r})$ has the lattice periodicity, $\mathbf{k}$ spans the first Brillouin zone, and $b$ is the band index. The eigenproblem $\hat{H}_0\psi_{b,\mathbf{k}} = \varepsilon_b(\mathbf{k}) \psi_{b,\mathbf{k}}$ is then solved, for given sets of $V_0$ and $\phig$, using exact numerical diagonalization in continuous space within the first Brillouin zone. It yields the numerically exact band dispersion relations~$\varepsilon_b(\mathbf{k})$
and the corresponding eigenstates~$\psi_{b,\mathbf{k}}(\mathbf{r})$.

Typical energy spectra are shown in Fig.~\ref{fig:bands} for various values of $V_0$ and $\phig$. \dng{For the balanced honeycomb lattice ($\phig=0$), the lowest band ($b=1$) consists of two subbands touching at two nonequivalent points of the first Brillouin zone (Dirac points $K_1$ and $K_2$), forming the characteristic Dirac cones with locally linear dispersion, see Figs.~\ref{fig:bands}(a), (b1), and (c1). Such Dirac cones are known to exist for almost any potential invariant under $2\pi/3$ rotation and possessing inversion symmetry ($\mathbf{r}\leftrightarrow-\mathbf{r}$)~\cite{Symmetry}.
They are also found in the second band ($b=2$). There, owing to the symmetries of the potential, each site hosts two degenerate first local excitations
associated with two Wannier states. For this reason, the second band splits into four subbands, corresponding to the two degenerate excited orbitals on each of the two sublattices. For finite geometric phase ($\phig>0$), the equivalence between the two sublattices is gradually broken, opening a gap at the Dirac points and progressively splitting both the lowest and excited bands, as shown in Figs.~\ref{fig:bands}(b2), (b3), (c2), and (c3). This gives rise to avoided crossings, altering the local dispersion from linear to quadratic\footnote{The avoided crossing is present in Fig.~\ref{fig:bands}(b2) but not visible due to the resolution.}.}

\dng{We now interpret these results in terms of TBMs  and assess their accuracy.} For deep enough potential, the lowest band can be understood from the TBM
corresponding to the first two lines in Eq.~(\ref{eq:TBM.Hamiltonian}), i.e.~for $\UA=\UB=0$.
The latter has been extensively studied previously, see for instance Refs.~\cite{Castro2009,Wallace1947,Javvaji,KeitaKishigi_2008}. 
\correct{The lowest band splits into two subbands ($\pm$) with dispersion relations} \dng{The dispersion relations of the two subbands ($\pm$) of the lowest band are}
\begin{eqnarray}
    \varepsilon_{1,\pm}\left(\mathbf{k}\right)
    &=&
    \pm J\sqrt{|f\left(\mathbf{k}\right)|^2 + \left[\frac{(\JA-\JB)f'(\mathbf{k})+\DeltaAB}{2J}\right]^2}
    \nonumber \\
    && - \frac{\JA+\JB}{2} f'(\mathbf{k}),
    \label{eq:TBMdispersion}
\end{eqnarray}
where $f(\textbf{k})=\sum_{j=1}^3e^{i\boldsymbol{\delta}_j\cdot\mathbf{k}}$ with $\boldsymbol{\delta}_j$ the lattice vectors connecting A-B pairs of nn sites
and
$f'(\textbf{k})=\sum_{j=1}^3e^{i\boldsymbol{\delta}'_j\cdot\mathbf{k}}$ with $\boldsymbol{\delta}'_j$ the lattice vectors connecting A-A or B-B pairs of nnn sites, see Fig.~\ref{fig:lattices}(b).
Explicit formulas are given in Appendix~\ref{app:TBM}. \dng{In the following, we compare three TBM descriptions of the lowest band. The full nnn model retains the four parameters $J$, $\JA$, $\JB$, and $\DeltaAB$ entering Eq.~\eqref{eq:TBMdispersion}. The nn model is obtained neglecting nnn terms, i.e., setting $\JA=\JB=0$. In the opposite limit, the triangular model is obtained by setting $J=0$, so that the A and B sublattices are described as two decoupled triangular lattices with tunneling amplitudes $\JA$ and $\JB$, respectively. See Appendix~\ref{app:TBM} for more details.}
The lowest band spectra we obtain for the continuous-space model are in good agreement with TBM predictions, Eq.~(\ref{eq:TBMdispersion}), for typically $V_0 \gtrsim 5\,\Er$.

More precisely, \correct{the BHm parameters $J$, $\JA$, $\JB$, and $\DeltaAB$ are extracted by fitting Eq.~(\ref{eq:TBMdispersion}) to the lowest-band dispersion relation obtained numerically from the continuous model} \dng{for each of the three models, the corresponding parameters are extracted by fitting its dispersion relation to the two lowest subbands obtained numerically from the continuous-space Hamiltonian}. Moreover, Wannier functions in A and B sites,
$\wA(\mathbf{r}-\mathbf{R}_j)$ and $\wB(\mathbf{r}-\mathbf{R}_j)$ with $\mathbf{R}_j$ the position of site $j$,
are obtained through a variational procedure that minimizes their real-space spread starting from the Bloch eigenstates~\cite{Wannier90}.
The on-site interaction energies, useful for the many-body problem, are then computed as
\begin{equation}
    \UAB = \tilde{g} \int d\mathbf{r}\, |\wAB(\mathbf{r})|^4.
\end{equation}

Finally, the quality of the fits is assessed through the error parameter, defined as the average of the absolute value of the difference between the TB dispersion and the exact dispersion, normalized by the corresponding subband width: 
\begin{equation}\label{eq:TBMerror}
    \epsilon_{\pm} = \frac{1}{\mathcal{A}\,S_\pm} \int d\mathbf{k}\ \left|\varepsilon_{1,\pm}\left(\mathbf{k}\right)-\varepsilon_{1,\pm}^{\textrm{\tiny TB}}\left(\mathbf{k}\right)\right|\text{,}
\end{equation}
where $\varepsilon_{1,\pm}\left(\mathbf{k}\right)$ and $\varepsilon_{1,\pm}^{\textrm{\tiny TB}}(\mathbf{k})$
are the dispersion relation, respectively,
obtained by diagonalizing the continuous Hamiltonian
and predicted by the TB model [Eq.~\eqref{eq:TBMdispersion} with fitted parameters],
$\mathcal{A}=\frac{\pi^2\sqrt3}{2a^2}$ is the area of the first Brillouin zone,
and $S_\pm$ is the bandwidth of the subband.
This procedure allows us to extract the effective tunneling parameters
and, at the same time, to assess the validity range of each model. 
Results for the lowest subbands are shown in Fig.~\ref{fig:fits}.\correct{for the full TBm as well as for some approximate models (see below)}

\correct{For $\phig=0$, we have $\DeltaAB=0$ and, for sufficiently deep lattices, the nnn terms can be neglected.
We then recover the standard dispersion relations of graphene, $\varepsilon_{1,\pm}\left(\mathbf{k}\right) \simeq \pm J|f(\mathbf{k})|$, which exhibit characteristic Dirac cones at $K_1$ and $K_2$ with locally linear dispersion relations, see Fig.~\ref{fig:bands}(a), (b1), and (c1). The latter are known to exist for almost any potential invariant under $2\pi/3$ rotation and possessing inversion symmetry ($\mathbf{r}\leftrightarrow-\mathbf{r}$); they are thus robust against the inclusion of nnn terms that respect such symmetry, as well as in continuous space~\cite{Symmetry}.
They are also found in the second band ($b=2$). There, owing to the symmetries of the potential, each site hosts two degenerate first local excitations
associated with two Wannier states. For this reason, the second band splits into four subbands, corresponding to the two degenerate excited orbitals on each of the two sublattices. In contrast, while the extreme subbands are strictly flat in the TBM restricted to nearest neighbors~\cite{PxPy}, we find weak $\mathbf{k}$ dependence in the continuous model for finite lattice amplitude $V_0$, see Fig.~\ref{fig:bands}(b1) and (c1).}\dng{For $\phig=0$, the two sublattices are equivalent, so that $\DeltaAB=0$ and $\JA=\JB\equiv J'$. For sufficiently deep lattices, the fitted nnn tunneling $J'$ is negligible compared with the nn tunneling $J$, and the full model is well approximated by the nn honeycomb model.
Consequently, the nn and full nnn fits yield close results in the balanced honeycomb limit, as shown in Figs.~\ref{fig:fits}(c) and (d). The higher bands require a different orbital description. In particular, a nn TBM constructed from the two degenerate first excited orbitals on each site predicts strictly flat extreme subbands~\cite{PxPy}. In the continuous-space spectra, however, these subbands retain a weak $\mathbf{k}$ dependence at finite lattice amplitude $V_0$, see Figs.~\ref{fig:bands}(b1) and (c1).}

%-------------------------------%
\begin{figure}[t]
    \centering
    \includegraphics[width=\linewidth]{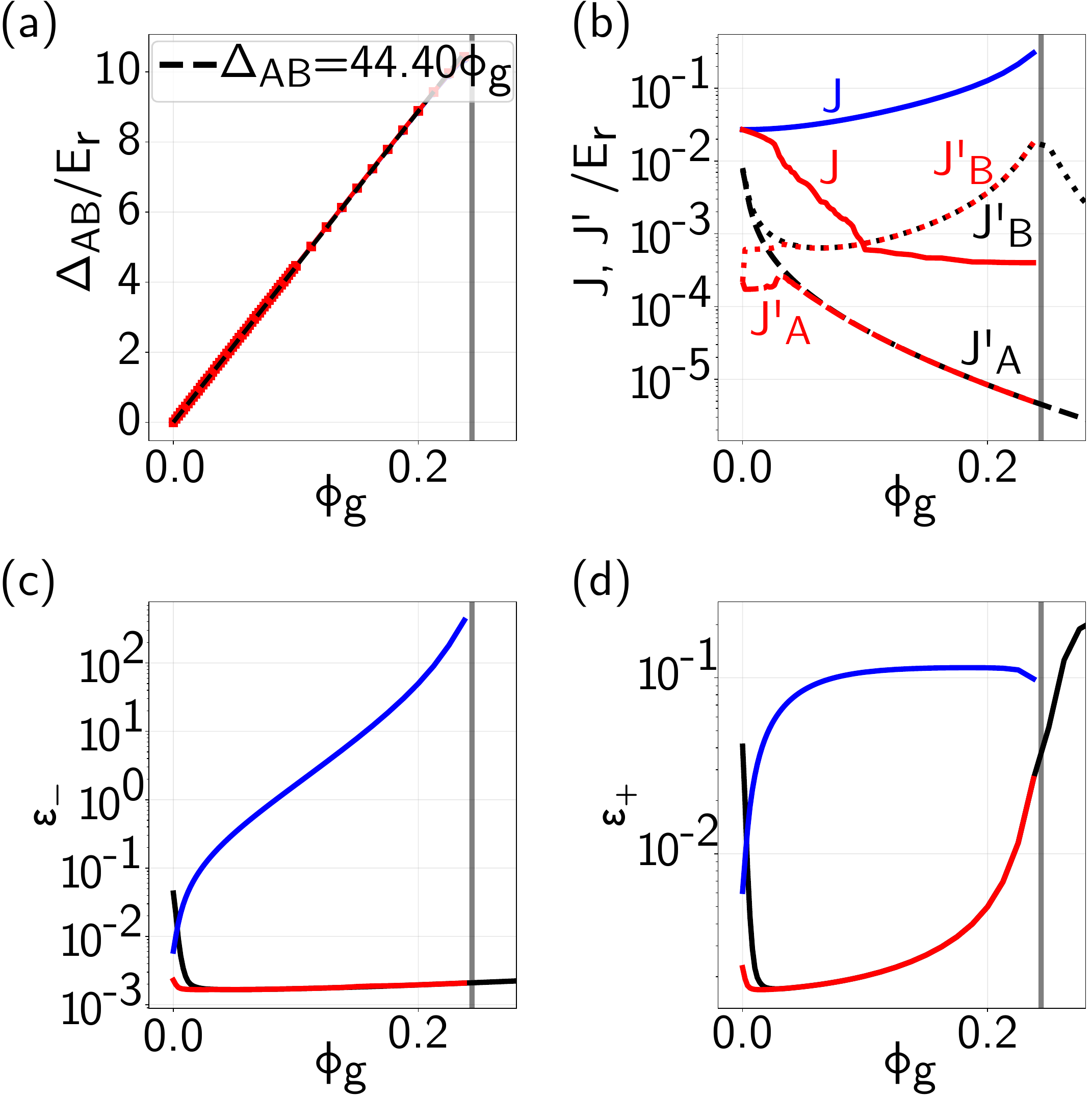}
    \caption{
Tight-binding parameters as functions of the geometric phase $\phig$ for $V_0=15\,\Er$.
Results obtained by fitting the lowest-band dispersion relation of the continuous single-particle Hamiltonian $\hat{H}_0$
to the nn~~(blue), nnn~~(red), and triangular~(black) models.
The vertical gray line marks $\phig^\ast\simeq0.25$, where the lowest B-site state crosses the first excited A-site state. 
(a)~Sublattice offset $\DeltaAB$ obtained from the nn (blue) and nnn (red) model (note that they overlap each other), together with a linear fit (dashed black line). 
(b)~Tunneling amplitudes: $J$ (solid), $\JA$ (dashed), and $\JB$ (dotted) for each model. 
(c) and (d)~Normalized fitting errors for, respectively, the lower and upper subbands.
}
    \label{fig:fits}
\end{figure}
%-------------------------------%

\correct{For finite~$\phig$, the equivalence of the sublattices A and~B is broken and a relative energy offset $\DeltaAB$ between A and B sites appears.
A band gap opens at the Dirac points, with local quadratic dispersion relations, while the bands and subbands progressively separate as $\phig$ (hence $\DeltaAB$) grows, see Fig.~\ref{fig:bands}(b2), (b3), (c2), and (c3).} \dng{For finite~$\phig$, the TBM description accounts for the broken equivalence of the $A$ and $B$ sublattices through a relative energy offset $\DeltaAB$.}
Expanding the energy difference between the ground states of the harmonic approximations of the potential around $A$ and $B$ sites, we find that $\DeltaAB$ grows linearly with $\phig$,
$
\DeltaAB \simeq (2\sqrt{3}V_0-\sqrt{3V_0\,\Er})\phig
$, see Appendix~\ref{app:Minima}.
This is in good agreement with fits to our numerical results, $\DeltaAB/\Er \simeq 44.40\,\phig$, see Fig.~\ref{fig:fits}(a).
The small discrepancy (less than $2\%$) may be attributed to the slight anharmonic effects.
\correct{Owing to the smaller separation between pairs of sites, the nn overlaps of Wannier functions always dominate their nnn counterparts.} For $\DeltaAB \ll 3J$, where $3J$ sets the half-bandwidth of the nn honeycomb dispersion [$\varepsilon_{1,\pm}(\mathbf{k})=\pm J|f(\mathbf{k})|$ and $|f(\mathbf{k})|\le 3$], we can rely on the nn model \correct{corresponding to neglecting the second line of Eq.~(\ref{eq:TBM.Hamiltonian}), and the primed terms in Eq.~(\ref{eq:TBMdispersion})}. The dispersion relations then reduce to $\varepsilon_{1,\pm}(\textbf{k}) \simeq \pm\frac{ \sqrt{\DeltaAB^2+4J^2|f(\textbf{k})|^2}}{2}$.
This is consistent with Fig.~\ref{fig:fits}, where we find that the nn model remains accurate for small $\phig$. A crossover occurs when $\DeltaAB\sim 3J$, which for $V_0=15\,\Er$ happens around $\phig\simeq 0.002$. The transition between effective descriptions is, however, gradual: in the vicinity of this crossover, the fitting errors of both the nn and triangular models remain reasonably small ($\lesssim 2\%$), so that either model provides a useful description of the low-energy physics. For larger $\phig$, nn tunneling becomes progressively off-resonant, while nnn tunneling remains resonant within each sublattice. The A and B sublattices then progressively decouple and the triangular-lattice description becomes more appropriate. The  terms in $J$ may be neglected and the dispersion relations, Eq.~(\ref{eq:TBMdispersion}), reduce to $\varepsilon_{1,\pm}\left(\mathbf{k}\right) \simeq \pm{\DeltaAB}/{2} \pm \JAB f'\left(\mathbf{k}\right)$ where $A$ corresponds to $-$ and $B$ corresponds to $+$. This is consistent with Fig.~\ref{fig:fits},
where we find that the tunnelings and errors obtained from the full (dotted and dashed red lines) and triangular (dotted and dashed black lines) models are similar for $\phig \gtrsim 0.025$, corresponding to $\DeltaAB/3J \gtrsim 20$ for $V_0=15\,\Er$. Finally, beyond a certain threshold, $\phig\!\gtrsim\!\phig^\ast$ (with $\phig^\ast\simeq0.25$ for $V_0=15\,\Er$, vertical lines in Fig.~\ref{fig:fits}), the energy offset $\DeltaAB$ exceeds the excitation energy on A sites, so that the ground-band Wannier state energy on B sites exceeds that of first excited states on A sites. Hence the only reasonable single-band TBM for the lowest subband  considered here becomes the triangular one for A sites.
Description of higher bands would involve including $p$-orbitals in A sites, $s$-orbitals in B sites, as well possibly their hybridation, which is beyond the scope of the present work.

\section{Quantum phase diagrams for interacting bosons}
We now turn to the many-body problem.
Low-temperature equilibrium properties of the full model of Eq.~\eqref{eq:Hamiltonian} for the potential of Eq.~\eqref{eq:potential} are found using \dng{continuous-space} worm-algorithm quantum PIMC calculations~\cite{wormalgo}.
We work within the grand-canonical ensemble with temperature $T$ and chemical potential $\mu$.
The interaction term is treated using the generalized propagator parametrized by the scattering length $\asc$, as introduced in Ref.~\cite{gautier2021}.
Unless stated otherwise, all simulations are performed at lattice depth $V_0=15\,\Er$ and temperature $T=0.01\,\Er/\kB$,
within in a $6\times 6$ unit-cell box with periodic boundary conditions, containing $72$ local lattice sites.
The observables extracted from our PIMC calculations are
the local particle density $\rho(\rr) = \langle \hat{\Psi}^\dagger(\rr) \hat{\Psi}(\rr) \rangle$,
the average particle number per site (i.e.~per half-unit cell)
$n=\int d\rr\, \rho(\rr)/N_c$, with $N_c$ the number of half-unit cells in the simulation box,
the average particle number in A and B sites, $n_A$ and $n_B$ [with $n=(n_A+n_B)/2$],
the compressibility $\kappa=\partial n/\partial\mu$,
and the SF fraction $f_s$.
The first five are computed using the particle position statistics while the latter is computed using the winding number estimator~\cite{ceperley1995}.
Our PIMC computations performed in continuous space and \dng{are interpreted in terms of and quantitatively compared to} the extended hBN-like BHM Hamiltonian of Eq.~(\ref{eq:TBM.Hamiltonian}).

\begin{figure*}[t!]
    \centering
    \includegraphics[width=\linewidth]{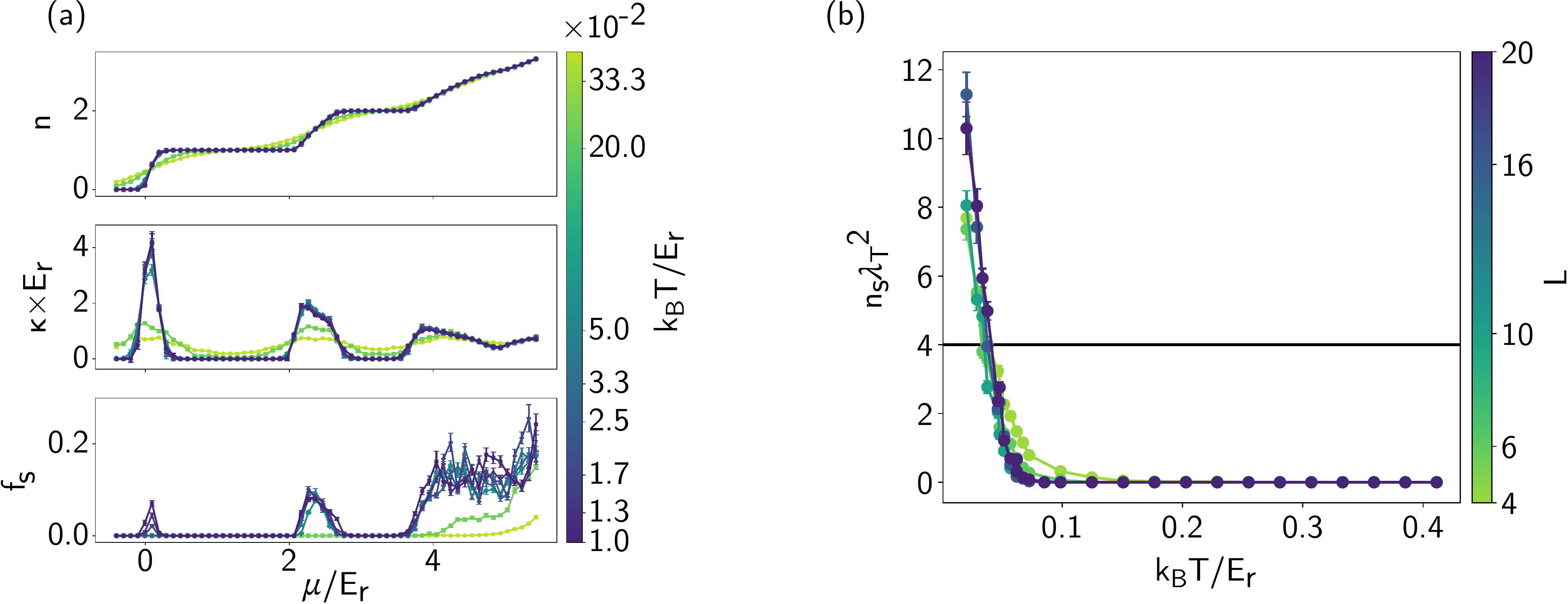}
    \caption{Quantum Monte Carlo simulation results for 2D bosons in a honeycomb optical lattice ($V_0=15\,\Er$, $\phig=0$, $\tgTwoD = 0.67$) for several temperatures and system sizes. 
~    (a)~Density $n$, compressibility $\kappa$, and SF fraction $\fs$ versus chemical potential for several temperatures (the ticks on the colorbar indicate the used temperatures). 
    (b)~Superfluid phase-space density $\ns\lambda_T^2$ versus temperature for various system sizes $L\times L$ (indicated on te colorbar), for $\mu \simeq 2.17\,\Er$. The black horizontal line indicates the expacted BKT SF transition.}
    \label{fig:temperature-size}
\end{figure*}

\subsection{Honeycomb lattice}
We start with the balanced honeycomb lattice, $\phig=0$. \dng{Typical PIMC results for the density per half unit cell $n$, the compressibility $\kappa$, and the SF fraction $\fs$ are shown in Fig.~\ref{fig:temperature-size}(a) for temperatures ranging from $T\simeq0.3\,\Er/\kB$ (lighter blue) to $T\simeq0.01\,\Er/\kB$ (darker blue). At sufficiently low temperatures, we clearly observe two MI plateaus at integer fillings $n=1$ and $n=2$, corresponding to one and two particles per lattice site, respectively. These plateaus are characterized by vanishing compressibility and SF fraction, and are separated by compressible SF regions with finite $\fs$. The system therefore undergoes successive SF--MI transitions as the chemical potential is increased.  Thermal fluctuations affect the insulating and SF properties on different energy scales. The density and compressibility remain only weakly modified for $\kB T\lesssim0.1\,\Er$, so that the MI plateaus remain well resolved over this temperature range. At higher temperatures, there is a progressive smoothing of the edges of the plateaus, eventually producing a finite compressibility within the regions that are insulating at low temperature. The SF response is more sensitive to temperature because it is controlled by the much smaller kinetic-energy scale. For the present lattice depth, the width of each lowest-band subband is approximately $0.08\,\Er$. Accordingly, when $\kB T$ becomes comparable to or larger than this bandwidth, phase coherence is strongly suppressed. For $\kB T\gtrsim0.2\,\Er$, the SF fraction vanishes throughout the chemical-potential range shown in Fig.~\ref{fig:temperature-size}(a), and the compressible regions correspond to normal-fluid (NF) phases. At intermediate temperatures, $0.05\,\Er\lesssim\kB T\lesssim0.2\,\Er$, the low-density SF phase is destroyed first, whereas the higher-density SF regions survive.}

\dng{More precisely, in 2D, the NF to SF transition is expected to be of the Berezinskii-Kosterlitz-Thouless (BKT) type~\cite{berezinskii1971,kosterlitz1973,nelson1977}. To check this, we study the SF phase-space density $\ns\lambdaT^2$ as a function of temperature $T$, where $\ns = \fs\times n$ is the SF density and $\lambdaT=\sqrt{2\pi\hbar^2/m\kB T}$ is the thermal de Broglie wavelength. A typical result for a fixed chemical potential $\mu$ is shown in Fig.~\ref{fig:temperature-size}(b) for various system sizes. The behavior shows a sharper and sharper transition as the size increases and a critical point consistent with the universal jump of the SF phase-space density at $\ns\lambdaT^2=4$, characteristics of the BKT transition.
Here we find a critical temperature of $\Tc \simeq 0.03\,\Er/\kB$.
Hence, the temperature used for the phase diagrams discussed below, $T = 0.01\,\Er/\kB$, is low enough to be representative of zero-temperature behavior.
Moreover, we find that, although the SF density is not fully converged, a system of $L\times L$ unit cells with $L=6$ is sufficient to locate the critical temperature clearly via the criterion $\ns\lambdaTc^2=4$.}

\begin{figure}[t]
    \centering
    \includegraphics[width=\linewidth]{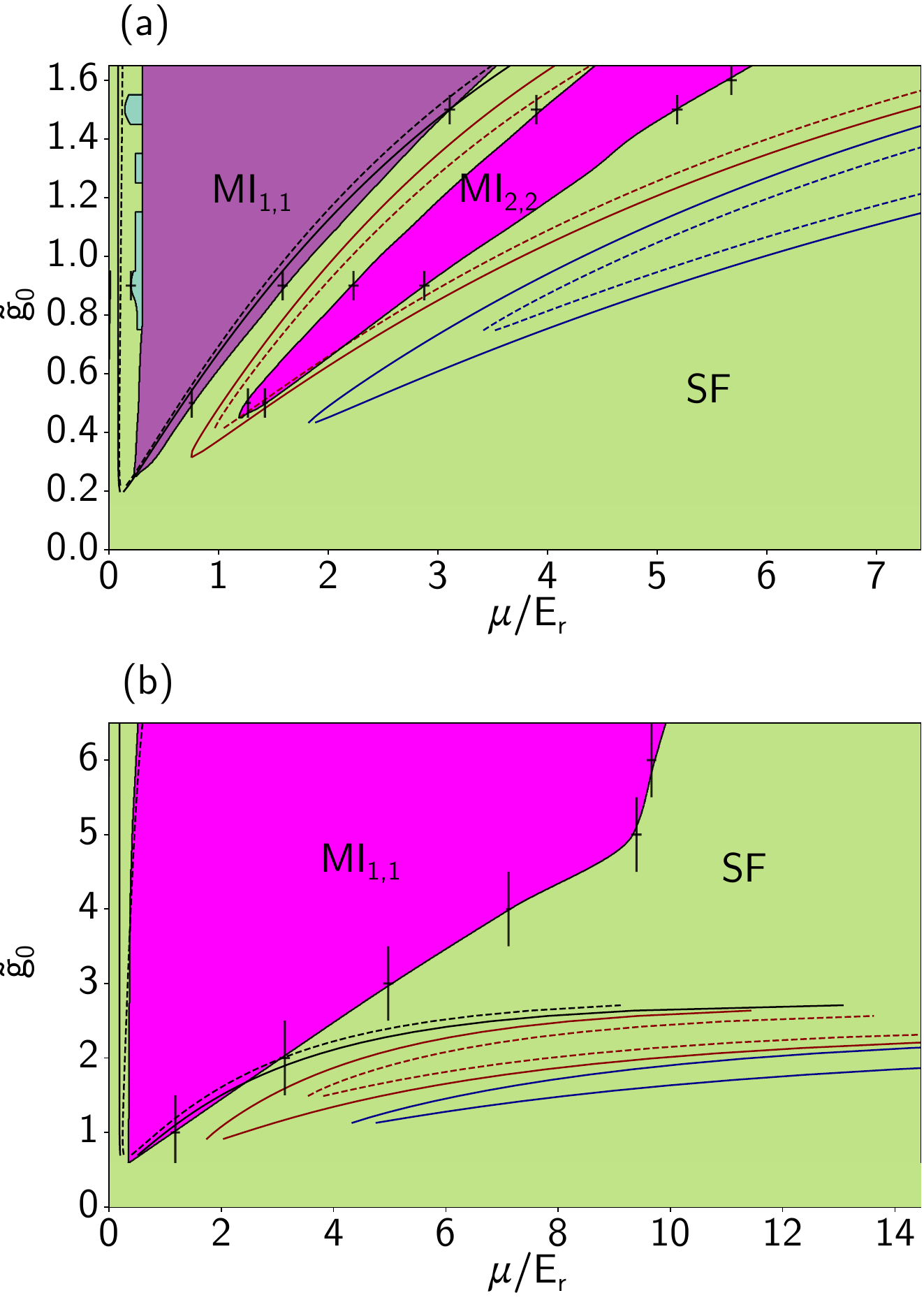}
    \caption{
Quantum phase diagram for bosons in the honeycomb lattice ($\phig=0$) versus chemical potential and interaction strength for (a)~$V_0=15\,\Er$ and (b)~$V_0=9.75\,\Er$
Results obtained using continuous-space PIMC for $\kB T=0.01\,\Er$,
Colored regions denote SF (light green), NF (dark green), and MI (purple-mauve) phases.
The crosses indicate selected boundary points obtained numerically with error bars corresponding to $\tgTwoD$--$\mu$ space discretization; only a subset is shown for visual clarity. Phase boundaries (solid black lines) are obtained by interpolation and extrapolation consistent with these error bars. Colored lines show Mott lobes predicted by the BHM, from Ref.~\cite{Teichmann2010}
using the nn tunneling extracted as in Fig.~\ref{fig:fits}(b) (solid lines) and including DAT corrections ($J\rightarrow J+2n\tilde J$, dashed lines).
}
    \label{fig:phasesphi0}
\end{figure}

\dng{The phase diagram of the balanced honeycomb model shown in Fig.~\ref{fig:phasesphi0}(a) is obtained by analyzing results as in Fig.~\ref{fig:temperature-size}(a) while varying the interaction strength $\tilde{g}_0$. At the used temperature, $T = 0.01\,\Er/\kB$, only very small regions exhibit finite compressibility and vanishing SF fraction, corresponding to a NF (dark green). The remainder of the diagram consists of two MI lobes with zero compressibility and zero SF fraction at integer fillings $n_A=n_B=1$ (MI$_{1,1}$, purple) and $n_A=n_B=2$ (MI$_{2,2}$, pink), surrounded by a broad SF phase (light green). The critical interaction strength for the appearance of the MI lobes is $\tgTwoD = 0.25 \pm 0.05$ for $V_0=15\,\Er$ as used in Fig.~\ref{fig:phasesphi0}(a).}

\dng{We now compare the continuous-space phase diagram with predictions of effective BHMs.} In the $\phig=0$ case, the two sublattices A and B are equivalent, $\DeltaAB=0$, $\UA=\UB\equiv U$, and $\JA=\JB\equiv J'$.
For $V_0 = 15\,\Er$, the single-particle results yield
$J\simeq2.7\times10^{-2}\,\Er$ and $J'\simeq2.2\times10^{-4}\,\Er$, and the nnn tunneling can be neglected.
The low-energy dynamics is thus essentially governed by the competition between onsite interactions $U$ and nn tunneling $J$ in Bose--Hubbard-like models.

\dng{The effective BHM parameters allow us to interpret the thermal behavior shown in Fig.~\ref{fig:temperature-size}(a). For $\tilde{g}_0\simeq0.67$, corresponding to $U\simeq1.83\,\Er$, the onsite interaction energy is approximately one order of magnitude larger than the temperatures for which the MI plateaus remain well resolved. This explains the weak temperature dependence of the density and compressibility for $\kB T\lesssim0.1\,\Er$. Moreover, the first MI gap at $n=1$ is approximately equal to $U$, as expected in the atomic limit $U\gg J$, whereas the second gap is significantly reduced owing to enhanced tunneling effects at $n\simeq2$ (see below). In contrast, the SF response is controlled by the much smaller tunneling energy scale. For low density, the relevant energy scale is the nn tunneling energy $J\simeq2.7\times10^{-2}\,\Er$, which is consistent with the onset of SF response observed in Fig.~\ref{fig:temperature-size}(a).}

The critical interaction strength found for the appearance of the MI lobes, $\tgTwoD = 0.25 \pm 0.05$, corresponds to $\Uc/J = 15.2 \pm 3.0$. This is comparable to BHM estimates for the honeycomb lattice, which yield
$\Uc/J \simeq 11.6$, obtained from high-order process-chain expansions~\cite{Teichmann2010}
and $\Uc/J \simeq 12.7$ from strong-coupling perturbation theory with series extrapolation~\cite{PhysRevA.85.023619}.
While our continuous-space phase diagram is qualitatively similar to BHM predictions (MI lobes corresponding to the estimate of Ref.~\cite{Teichmann2010} are shown as thin dashed lines), it exhibits clear quantitative deviations. The MI lobe at $n_A=n_B=1$ agrees well up to an approximate shift of the chemical potential of order $\Delta\mu=0.15\,\Er$. In contrast, the MI lobe at $n_A=n_B=2$ exhibits a gap that is approximately half the size predicted by the BHM. Furthermore, the MI lobe at $n_A=n_B=3$ predicted by the BHM is completely absent from the full continuous-space PIMC calculations. In fact, the PIMC results show a weak inflection point of the $n(\mu)$ curve at $n=3$,
but the compressibility and the SF fraction remain finite, see Fig.~\ref{fig:temperature-size}(a). This indicates enhanced correlations, but not strong enough to stabilize a MI phase.
These deviations, in particular the suppression of the $n=2$ lobe and the absence of the $n=3$ lobe,
indicate that the BHM fails to completely capture the low-energy physics in this regime,
highlighting the need for continuous-space calculations.

To explain these discrepancies, we may consider corrections to the standard BHM.
We first note that the higher single-particle bands, which appear at energies around $8.75\,\Er$ for $V_0=15\,\Er$, see Fig.~\ref{fig:bands}(c1), lie well above the chemical-potential range considered in the phase diagram of Fig.~\ref{fig:phasesphi0}(a) and can therefore be ignored. 
We may then consider nnn tunneling terms, corresponding to the second line of Eq.~\eqref{eq:TBM.Hamiltonian}, as well as nn interactions, which generate a term of the form $\frac{U'}{2}\sum_{\langle j,\ell \rangle} \hat{n}_j\hat{n}_\ell$,
where
$U'=\tilde{g} \int d\rr\, |\wA|^2|\wB|^2$.
Such terms are expected to play a role for shallow lattices owing to enhanced spatial extension of the Wannier functions.
However, for $V_0=15\,\Er$ as used here our single-particle calculations yield
$J' \simeq 2.18\times10^{-4}\,\Er$ and $U'/\tilde{g}\simeq1.9\times10^{-4}\,\Er$.
These terms are, respectively, about two and four orders of magnitude smaller than the nn tunneling ($J \simeq 2.7\times10^{-2}\,\Er$) and on-site interaction strength ($U/\tilde{g}  \simeq 2.73\,\Er$), and can safely be neglected.
In contrast, density-assisted tunneling (DAT), which arises from nn coupling terms,
$-\tilde{J}\sum_{\langle j,\ell \rangle} \hat{a}_j^\dagger(\hat{n}_j+\hat{n}_\ell)\hat{a}_\ell$ with $\tilde{J}=-\tilde{g} \int d\mathbf{r}\, |\wA|^2 \wA^* \wB$, becomes significant.
Such terms are beyond BHM. Nevertheless, we may estimate their significance within a mean-field approach by replacing the density operator $\hat{n}_j+\hat{n}_\ell$ by its average value $2n$. We then recover the BHM where the hopping energy $J$ must be replaced by the effective hopping energy $J+2n\tilde{J}$. Here, $\tilde{J}$ is evaluated using the bare coupling $\tilde g_0$; since in most of the relevant parameter range $\tilde g>\tilde g_0$, this provides a conservative estimate of DAT effects. Examination of this term shows that the mean-field DAT correction is on the order of $30\%$ near the center of the Mott lobe at $n=1$, on the order of $70\%$ in the lobe at $n=2$, and on the order of $100\%$ in the lobe at $n=3$, see Appendix~\ref{app:DAT}. The Mott lobes estimated from BHM predictions and including the mean-field DAT corrections are shown as dashed lines in the phase diagrams of Fig.~\ref{fig:phasesphi0}. These results show that, while the MI lobe at $n=1$ is marginally affected, the lobes at $n=2$ and $n=3$ are strongly reduced. This suggests that DAT terms are important, even for a large lattice depth such as $V_0=15\,\Er$. The simple mean-field treatment used here is qualitatively consistent\correct{with our PIMC results} \dng{with the skrinking of Mott lobes with $n>1$ observed in our continuous-space PIMC calculations.} Quantitative agreement, particularly the complete disappearance of the lobe at $n=3$, would, however, require a \correct{full quantum} \dng{beyond mean-field} treatment of the DAT terms.\correct{, which is beyond the BHM.}

As expected, the breakdown of the BHM becomes more pronounced for shallower lattices. 
Fig.~\ref{fig:phasesphi0}(b) shows the phase diagram for the same system at $V_0=9.75\,\Er$. 
Qualitatively, the situation remains similar to the case at $V_0=15\,\Er$, but with significantly stronger deviations from the BHM description. First, density-assisted tunneling (DAT) effects are enhanced in shallower lattices due to the increased spatial extent of the Wannier functions, leading to a larger relative correction to the bare tunneling amplitude. 
Second, the relevant chemical potentials now extend beyond the gap separating the lowest band from the first excited band. In particular, the first Mott lobe is stabilized at chemical potentials exceeding this gap, signaling that higher-band processes already play a role at unit filling.
Third, the effective interaction description based on Eq.~\eqref{eq:gRenorm} is no longer accurate in this regime. That expression relies on the assumption of sufficiently small $\tilde{g}_0$ and $\mu$, whereas the parameters considered here lie outside this perturbative regime.
Taken together, these three effects lead to substantial deviations from the single-band BHM at $V_0=9.75\,\Er$.

\subsection{Hexagonal boron nitride lattice}
\begin{figure}[!t]
    \centering
    \includegraphics[width=\linewidth]{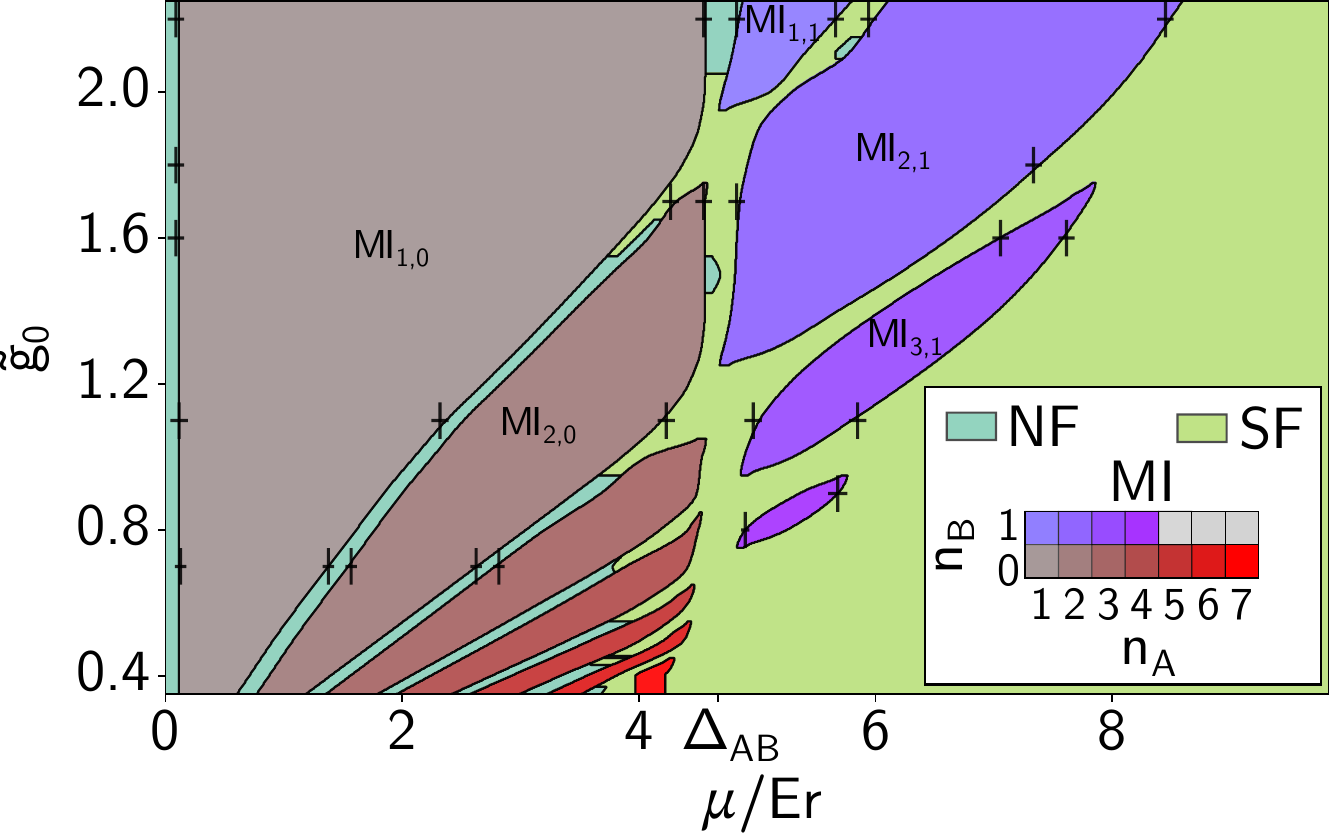}
    \caption{
    Quantum phase diagram for bosons in the asymmetric h-BN lattice with $\phig=\pi/30$ and $V_0=15\,\Er$ (corresponding to $\DeltaAB\simeq4.7\,\Er$) versus chemical potential and interaction strength.
    Results obtained using continuous-space PIMC for $\kB T=0.01\,\Er$. Colored regions denote SF (light green), NF (dark green), and MI (red, brown, purple; labeled by the sublattice occupations $MI_{n_A,n_B}$) phases. The crosses indicate selected boundary points obtained numerically with error bars corresponding to $\tgTwoD$--$\mu$ space discretization; only a subset is shown for visual clarity. Phase boundaries (solid black lines) are obtained by interpolation and extrapolation consistent with these error bars.
    }

    \label{fig:phipi30}
\end{figure}

We finally consider the asymmetric case, where $\phig \neq 0$.
This case was studied for unit filling, $n_A+n_B=2$, in Ref.~\cite{wang2025},
where it was shown that competition between the energy offset $\DeltaAB$ and on-site interaction energy induces
transitions from SF to MI$_{1,1}$ and MI$_{2,0}$ phases.
Beyond unit filling, we find a very rich phase diagram, shown in Fig.~\ref{fig:phipi30} for
$V_0=15\,\Er$ and $\phig=\pi/30$, corresponding to $\DeltaAB \simeq 4.7\,\Er$,
and $\kB T \simeq0.01\,\Er$.
Each colored region represents a SF (light green), a NF (dark green), and distinct Mott phases, labeled by the integer
occupations $(n_A,n_B)$ of the two sublattices. \dng{At low interaction strengths, particles preferentially occupy the energetically favorable A sublattice, giving rise to a sequence of insulating phases with $n_B=0$ as $\mu$ increases. For $\mu > \DeltaAB$, the B sites are populated but strong tunneling to A sites gives rise to a SF phase.
For large-enough interaction strength, multiple occupancy of the A sites becomes energetically costly so that the MI lobes with low $n_A$ and $n_B=0$ get larger. Moreover, nn tunneling is overcome by interactions and, for $\mu > \DeltaAB$, particles progressively populate the B sites, leading to the emergence of Mott phases with finite occupation on both sublattices. Depending on the relevant tunneling processes, these insulating phases are separated either by broad SF regions or by extremely narrow NF or SF regions. Altogether, the phase diagram reveals a rich interplay between interaction effects and sublattice imbalance that has no counterpart in the balanced honeycomb case. Note that the results of Ref.~\cite{wang2025} were obtained within a BHM description at unit filling, whereas the present phase diagram is obtained directly from continuous-space PIMC simulations and extends well beyond the unit-filling regime.}

The overall structure of the phase diagram in Fig.~\ref{fig:phipi30} can be qualitatively understood
from the atomic limit ($J,\JA,\JB\!\to\!0$) of the extended BHM of Eq.~\eqref{eq:TBM.Hamiltonian}.
For $\UA\ll\DeltaAB$ and $\mu < \DeltaAB$, only the A sites are occupied and
a series of $(n_A,0)$ Mott plateaus appear as $\mu$ increases until $n_A\UA\simeq\DeltaAB$ or $n_A$ is too large to stabilize a MI against Bose-enhanced tunneling. In both cases, the B sites are completely irrelevant for the MI phases. Because the relevant tunneling in this regime is the small nnn term $\JA$, the SF windows between MI plateaus are extremely narrow. Moreover, superfluidity would require a temperature typically smaller than $\JA$, which is about two orders of magnitude smaller than our working temperature and we consistently find a NF phase.
This behavior is observed in the phase diagram for \correct{$\tgTwoD\!\in\![0.24,0.43]$} \dng{$\tgTwoD\!\in\![0.39,0.70]$}, where the system effectively behaves
as a weakly coupled triangular lattice on the A sublattice.

As $\tgTwoD$ increases, the onsite interaction $\UA$ grows and the condition
$\DeltaAB \simeq n_A\UA$ is reached.
At this point, the B sites start to populate, giving rise to the
sequence of Mott lobes $(n_A,0)\ \rightarrow\ (n_A{+}1,1)$,
accompanied by broad SF regions between the Mott lobes.
This regime is clearly visible in the ranges
\correct{$\Tilde{g}_0\!\in\![0.49,0.55]$}\dng{$\Tilde{g}_0\!\in\![0.80,0.90]$}, \correct{$\Tilde{g}_0\!\in\![0.67,0.73]$} \dng{$\Tilde{g}_0\!\in\![1.10,1.20]$}, and \correct{$\Tilde{g}_0\!\in\![1.09,1.16]$} \dng{$\Tilde{g}_0\!\in\![1.79,1.91]$}.
When $(n_A{-}1)\UA < \DeltaAB < n_A\UA$,
the sequence changes to $(n_A,0)\!\rightarrow\!(n_A,1)$,
as observed for \correct{$\tgTwoD \simeq 0.61$ and $\tgTwoD \in [0.73,1.03]$} \dng{$\tgTwoD \simeq 1.00$ and $\tgTwoD \in [1.20,1.69]$}.
The behavior discussed in Ref.~\cite{wang2025} for unit filling, corresponds to two transitions from MI$_{2,0}$ to SF and from SF to MI$_{1,1}$ observed in our diagram for $\mu \simeq \DeltaAB$ and \correct{$\tgTwoD\in [1.06,1.18]$} \dng{$\tgTwoD\in [1.74,1.94]$}.
Note that for $\mu \gtrsim \DeltaAB$, the relevant tunnelings is $J$ corresponding to hopping between nn A and B sites.
It is larger than our working temperature and we consistently find SF regions instead of NF regions.

\section{Conclusions}
In this work, we have investigated the dynamics of ultracold bosons in honeycomb and h-BN-like optical potentials, combining band structure calculations with exact PIMC simulations, both in continuous-space. Our results reveal complex phase diagrams, which may significantly deviate from the predictions of standard BHMs, even in regimes where single-particle properties are accurately described by TBM approximations. For the honeycomb lattice, \correct{we find} \dng{our mean-field estimate suggests} that density-assisted tunneling, beyond the BHM frameworks, plays a crucial role suppressing higher-order MI lobes and reshaping the phase boundaries\dng{, but a full assessment would require a beyond-mean-field treatment}. In the asymmetric h-BN lattice, the interplay between sublattice energy offset, tunneling, and interactions gives rise to a cascade of MI lobes with a variety of sublattice occupations, as well as broad SF domains.
These findings underscore the importance of continuous-space approaches for accurately capturing the many-body physics of bosons in hexagonal lattices, particularly in regimes where beyond-nn tunneling and interaction-induced corrections become significant.
These conclusions are in line with previous work in other nonstandard geometries, such as quasicrystal~\cite{gautier2021,ciardi2022,zhu2023} and twisted potentials~\cite{johnstone2024}.
The rich phenomenology uncovered here invites further theoretical and experimental investigations, with potential implications for quantum simulation and the design of novel synthetic materials in hexagonal geometries.
Our work not only provides a guide for future experiments with ultracold atoms or cavity polaritons in hexagonal potentials but also opens new avenues for exploring exotic quantum matter in other configurations, including superlattices, frustrated models, and multilayer systems.

%%%%%%%%%%%%%%%%%%%%%%%%%%%%%%
\begin{acknowledgments}
We thank Yifei Wang for early-stage contributions to the formulation of the single-particle problem.
This work was supported by the Agence Nationale de la Recherche under projects QuanTEdu-France (ANR-CMAQ-002 France~2030) and QUTISYM (ANR-23-PETQ-0002),
and by computing HPC and storage resources by GENCI at TGCC under the grants AD010510300R2 and AP010510288.
\end{acknowledgments}

%%%%%%%%%%%%%%%%%%%%%%%%%%%%%%%%%%%%%%%%%%%%%%%%%%%%%%%%%%%%
%\clearpage
\appendix

%%%%%%%%%%%%%%%%%%%%%%%%%%%%%%
\section{Geometry of the minima as a function of the geometric phase}
\label{app:Minima}

In this Appendix, we analyze the extrema of the lattice potential as a function of the geometric phase $\phig$. This determines the positions of the lattice sites used in the main text, the range of $\phig$ for which the potential contains two minima per unit cell, and the point at which only one minimum per lattice site is left and the lattice geometry reduces to a triangular one. We also derive the small-$\phig$ expression for the sublattice offset $\DeltaAB$.

We consider the potential of Eq.~\eqref{eq:potential} with $\phi_1=\phi_2=\phi_3=\phig/3$. As in Fig.~\ref{fig:lattices}(a) of the main text, we orient the reciprocal vectors as
\begin{equation}
\begin{cases}
\mathbf G_1 = -\dfrac{\pi}{a}\mathbf e_y,\\
\mathbf G_2 = \dfrac{\pi}{a}\left(\dfrac{\sqrt{3}}{2}\mathbf e_x + \dfrac{1}{2}\mathbf e_y\right),\\
\mathbf G_3 = \dfrac{\pi}{a}\left(-\dfrac{\sqrt{3}}{2}\mathbf e_x + \dfrac{1}{2}\mathbf e_y\right).
\end{cases}
\end{equation}

Introducing the variables
\begin{equation}
\theta_j=\mathbf G_j\!\cdot\!\mathbf r\label{eq:app-theta},
\end{equation}
and using $\mathbf G_1+\mathbf G_2+\mathbf G_3=0$, the potential depends on two independent variables only,
\begin{align}
V(\theta_1,\theta_2)=2V_0\Big[
&\cos(\theta_1+\phig/3)+\cos(\theta_2+\phig/3)
\nonumber\\
&+\cos(\theta_1+\theta_2-\phig/3)
\Big].
\label{eq:app-potential}
\end{align}
It is sufficient to restrict the analysis to one unit cell, $\theta_{1,2}\in[0,2\pi)$ and take $\phig\in(-\pi,\pi]$.

\subsection{Stationary points and classification}

The potential extrema satisfy
\begin{equation}
\begin{cases}
    \partial_{\tu}V \propto\sin(\tu+\phig/3)+\sin(\tu+\td-\phig/3)=0, \\
    \partial_{\td}V \propto \sin(\td+\phig/3)+\sin(\tu+\td-\phig/3)=0.
\end{cases}
\label{eq:app-stationary}
\end{equation}
Subtracting the two equations gives
\begin{equation}
\sin(\theta_1+\phig/3)=\sin(\theta_2+\phig/3),
\end{equation}
which yields two branches of stationary points:
\begin{equation}
\begin{cases}
    \text{Branch A: }\theta_1=\theta_2 \\
    \text{Branch B: }\theta_2=\pi-\theta_1.
\end{cases}
\end{equation}

For Branch A, setting $\theta_1=\theta_2\equiv\theta$, the stationarity condition~\ref{eq:app-stationary} becomes
\begin{equation}
\sin(\theta+\phig/3)+\sin(2\theta-\phig/3)=0.
\end{equation}
Using trigonometric identities, this yields two families of solutions:
\begin{equation}
\theta=\frac{2\pi n}{3}
\quad \text{with} \quad n=0,1,2,\label{eq:app-branch1}
\end{equation}
and
\begin{equation}
\theta=\pi+\frac{2\phig}{3}\label{eq:app-branch2}.
\end{equation}

For Branch B, the solutions are
\begin{equation}
(\theta_1,\theta_2)=(\pi,0),
\qquad
\left(2\pi-\frac{2\phig}{3},\frac{2\phig}{3}-\pi\right).
\end{equation}

We now determine the nature of the stationary points from the Hessian matrix
\begin{equation}
H_{j\ell}=\frac{\partial^2V}{\partial\theta_j\partial\theta_\ell}.
\end{equation}

For all Branch B solutions, as well as for the Branch A solution of Eq.~\eqref{eq:app-branch2}
one finds $\det H\le0$, so that these points are saddle points.

For the three Branch A solutions of Eq.~\eqref{eq:app-branch1}
the Hessian matrix reads
\begin{equation}
H=-2V_0
\begin{pmatrix}
2c_n & c_n\\
c_n & 2c_n
\end{pmatrix},
\quad
c_n=\cos\!\left(\frac{\phig}{3}+\frac{2\pi n}{3}\right).
\end{equation}
Hence, we find
\begin{equation}
\det H=12V_0^2c_n^2\ge0
\quad\text{and}\quad
\mathrm{Tr}\,H=-12V_0c_n,
\end{equation}
so that these points are minima when $c_n<0$ and maxima when $c_n>0$. This implies that the stationary point with $n=0$ is always a maxima for any value of $\phig$ in our considered range and only $n=1$ and $n=2$ rest as candidate minima. In summary, among all stationary points, only those of the Branch A given by Eq.~\eqref{eq:app-branch1} with $n=1$ or $n=2$ can be minima.

\subsection{Number and position of the minima}

The sign of
\begin{equation}
c_n=\cos\!\left(\frac{\phig}{3}+\frac{2\pi n}{3}\right)
\end{equation}
determines how many minima are present.

\begin{itemize}
\item For $|\phig|<\pi/2$, both $n=1$ and $n=2$ are minima. The lattice therefore contains two potential minima (lattice sites) per unit cell. For $\phig=0$, the two sites are degenerate in energy and form the balanced honeycomb lattice. For $\phig\neq0$, the degeneracy is lifted, yielding the imbalanced geometry discussed in the main text.

\item For $|\phig|>\pi/2$, only one minimum remains corresponding to $n=1$ for $\phig>0$ and to $n=2$ for $\phig<0$. The lattice then reduces to a triangular lattice.
\end{itemize}

To find the positions of the minima, we solve Eq.~\eqref{eq:app-theta}
\begin{equation}
\theta_1=\theta_2=\frac{2\pi n}{3}\quad\text{with}\quad\begin{cases}
    n=1\\n=2\text{ .}
\end{cases}
\end{equation}
With the orientation of Fig.~\ref{fig:lattices}(a) of the main text, we find the real-space positions
\begin{equation}
\mathbf r_1=
\frac{2a}{\sqrt3}\mathbf e_x-\frac{2a}{3}\mathbf e_y,
\qquad
\mathbf r_2=
\frac{4a}{\sqrt3}\mathbf e_x-\frac{4a}{3}\mathbf e_y.
\end{equation}
Remarkably, these positions are independent of the geometric phase for the choice $\phi_1=\phi_2=\phi_3=\phig/3$.

\begin{figure}[t]
\centering
\includegraphics[width=\columnwidth]{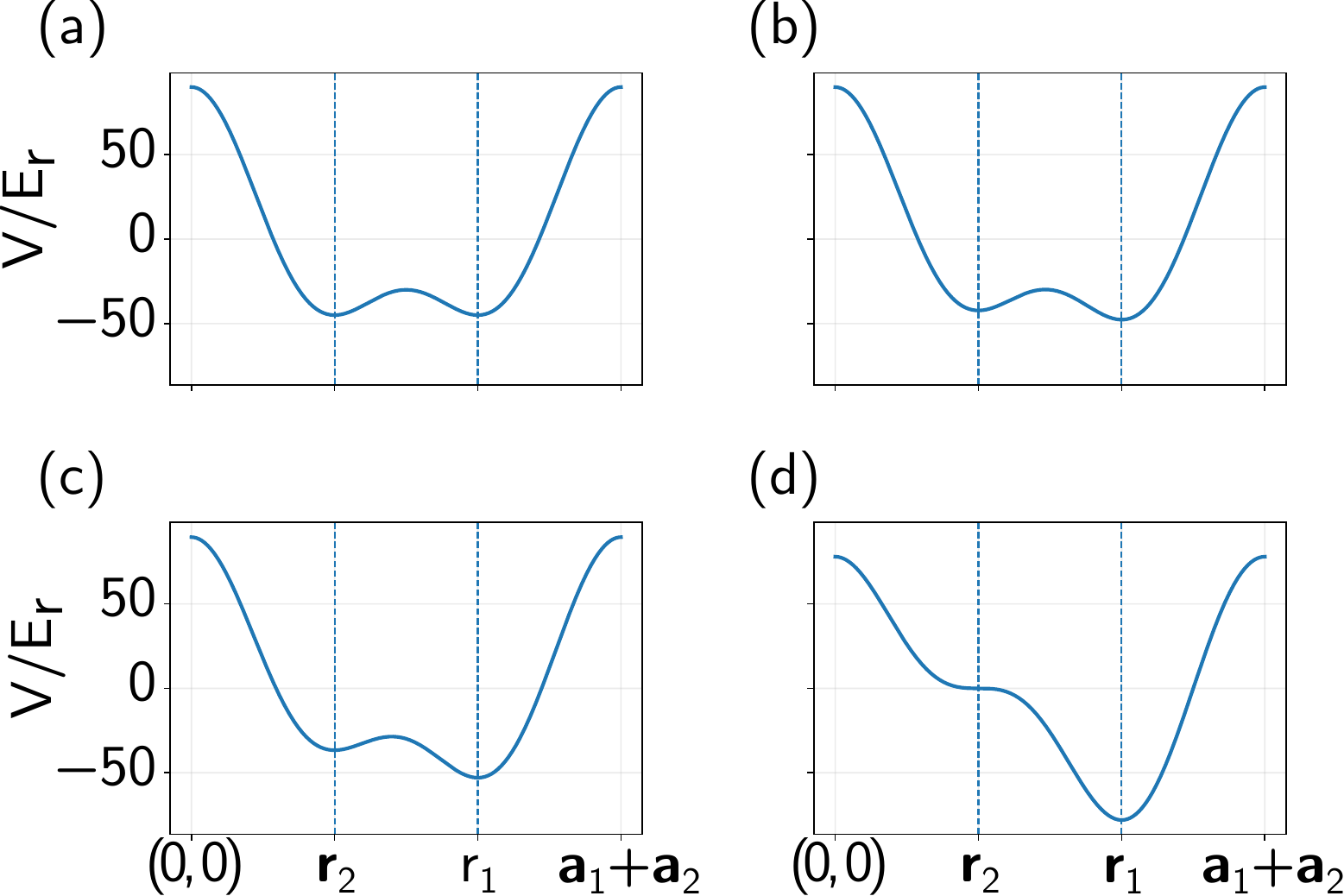}
\caption{
Line cut of the lattice potential at $V_0=15\,\Er$ along the path
$(0,0)\rightarrow \mathbf r_2 \rightarrow \mathbf r_1 \rightarrow
\mathbf a_1+\mathbf a_2$, for
(a) $\phig=0$,
(b) $\phig=\pi/30$,
(c) $\phig=\pi/10$,
and (d) $\phig=\pi/2$.
The dashed vertical lines mark the positions of $\mathbf r_2$ and $\mathbf r_1$.}
\label{fig:app_linecut}
\end{figure}

For $\phig>0$, the $n=1$ site is lower in energy and is identified with the A sublattice, while the $n=2$ site corresponds to the B sublattice. For $\phig<0$, the roles are exchanged. To visualize the evolution of these extrema, Fig.~\ref{fig:app_linecut}
shows a one-dimensional cut of the potential taken along the line
connecting $(0,0)$, $\mathbf r_2$, $\mathbf r_1$, and $\mathbf a_1+\mathbf a_2$.
For $\phig=0$, the two minima are degenerate, see Fig.~\ref{fig:app_linecut}(a). Increasing to $\phig=\pi/30$ and to $\phig=\pi/10$
lifts this degeneracy, see Figs.~\ref{fig:app_linecut}(b) and (c). Finally, for $|\phig|=\pi/2$ one of the minima
disappears, consistently with the transition to a triangular lattice, see Fig.~\ref{fig:app_linecut}(d).

\subsection{Energy splitting and sublattice offset}

We now focus on the regime $|\phig|<\pi/2$, where the both extrema $n=1,2$ are local minima. The offset $\DeltaAB$ between the two sublattices has two contributions. The first one comes from the different values of the potential at the two minima. The second one comes from the different local curvatures of the wells, which yield different ground-state energies for the corresponding local harmonic oscillators.

The potential energy at each of the two minima is given by Eq.~\eqref{eq:app-potential} with
\begin{equation}
\theta_1=\theta_2=\frac{2\pi n}{3}\quad \text{and}
\quad n=1,2,
\end{equation}
which yields
\begin{equation}
V_n
=
6V_0\cos\!\left(\frac{\phig}{3}+\frac{2\pi n}{3}\right).
\end{equation}
The classical contribution to the energy splitting is thus
\begin{align}
\Delta V
&=V_2-V_1
\nonumber\\
&=
6V_0\left[
\cos\!\left(\frac{\phig}{3}+\frac{4\pi}{3}\right)
-
\cos\!\left(\frac{\phig}{3}+\frac{2\pi}{3}\right)
\right].
\end{align}
Using trigonometric identities, we then find
\begin{equation}
\Delta V
=
6\sqrt3\,V_0\sin\!\left(\frac{\phig}{3}\right).
\end{equation}
For $|\phig|\ll1$, it yields
\begin{equation}
\Delta V\simeq2\sqrt3\,V_0\,\phig.
\label{eq:app-DeltaV}
\end{equation}
This is the splitting between the local potential minima.

To find the energy offset, we now include the harmonic zero-point contribution. Expanding the potential to quadratic order around a minimum $\mathbf r_n$ gives
\begin{equation}
V(\mathbf r)
\simeq
V_n
+
\frac12
\sum_{\alpha,\beta=x,y}
H^{(n)}_{\alpha\beta}
\delta r_\alpha \delta r_\beta,
\end{equation}
where
\begin{equation}
    H^{(n)}_{\alpha\beta}=\left.\frac{\partial^2V}{\partial\alpha\partial\beta}\right|_{\rr=\rr_n}
\end{equation}
is the Cartesian Hessian matrix.
For the minima considered here, the Cartesian Hessian is diagonal,
\begin{equation}
H^{(n)}_{\alpha\beta}
=
-\frac{3\pi^2}{a^2}V_0 c_n\,\delta_{\alpha\beta}.
\end{equation}
The local harmonic frequency such that $V(\rr_n+\delta\rr)\approx V_n+m\omega_n^2(\delta\rr)^2/2$ is therefore
\begin{equation}
\omega_n^2=
-\frac{3\pi^2}{ma^2}V_0 c_n.
\end{equation}
Using the recoil-energy convention of the main text, $\Er={\pi^2\hbar^2}/{6ma^2}$,
this may be written as
\begin{equation}
\omega_n
=
\frac{1}{\hbar}
\sqrt{18V_0\Er(-c_n)}\text{ }.
\end{equation}
Since the local harmonic oscillator is two-dimensional and isotropic, the local zero point (zp) energy above the potential minimum $V_n$ is
$E_{\textrm{zp}}^{(n)} = \hbar\omega_n$, that is,
\begin{equation}
E_{\textrm{zp}}^{(n)}
=
\sqrt{18V_0\Er}\sqrt{-c_n}.
\end{equation}

Expanding for $\phig\ll1$, we find
\begin{equation}
c_{1,2}
=
-\frac12
\mp
\frac{\sqrt3}{6}\phig
+
O(\phig^2),
\end{equation}
where the upper sign corresponds to $n=1$ and the lower sign to $n=2$, and we obtain
\begin{equation}
\sqrt{-c_2}-\sqrt{-c_1}
=
-\frac{\sqrt3}{3\sqrt2}\phig
+
O(\phig^2).
\end{equation}
The zero-point contribution to the splitting is therefore
\begin{equation}
E_{\textrm{zp}}^{(2)}-E_{\textrm{zp}}^{(1)}
=
-\sqrt{3V_0\Er}\,\phig
+
O(\phig^2).
\label{eq:app-DeltaEzp}
\end{equation}

Combining Eqs.~\eqref{eq:app-DeltaV} and \eqref{eq:app-DeltaEzp}, the energy difference between the two lowest local states in A and B sites is thus
\begin{equation}
\DeltaAB
=
\left(
2\sqrt3\,V_0
-
\sqrt{3V_0\Er}
\right)\phig\text{ ,}
\end{equation}
to lowest order in $\phig$.

%%%%%%%%%%%%%%%%%%%%%%%%%%%%%%
\section{Tight-binding models}\label{app:TBM}

In this Appendix, we give the explicit TBM conventions used in the main text. We first specify the lattice vectors and hopping geometry, then write the corresponding single-particle tight-binding Hamiltonian and dispersion relation. We finally give the nn and triangular limits used to analyze and fit band structures.

\subsection{Lattice vectors and hopping geometry}

We recall the Bravais vectors used in the main text,
\begin{equation}
    \mathbf{a}_1=\frac{4a}{\sqrt{3}}\,\mathbf{e}_x,\qquad
    \mathbf{a}_2=\frac{2a}{\sqrt{3}}\,\mathbf{e}_x+2a\,\mathbf{e}_y .
\end{equation}
The nn vectors connecting an A site to the three neighboring B sites, see Fig.~\ref{fig:lattices}(b), are
\begin{equation}
\begin{cases}
    \boldsymbol{\delta}_1
    &=
    \frac{-\mathbf{a}_1+2\mathbf{a}_2}{3}
    =
    \frac{4a}{3}\,\mathbf{e}_y,
    \\
    \boldsymbol{\delta}_2
    &=
    \frac{2\mathbf{a}_1-\mathbf{a}_2}{3}
    =
    \frac{2a\sqrt{3}}{3}\,\mathbf{e}_x
    -\frac{2a}{3}\,\mathbf{e}_y,
    \\
    \boldsymbol{\delta}_3
    &=
    -\frac{\mathbf{a}_1+\mathbf{a}_2}{3}
    =
    -\frac{2a\sqrt{3}}{3}\,\mathbf{e}_x
    -\frac{2a}{3}\,\mathbf{e}_y .
\end{cases}
\end{equation}
The nnn vectors connect sites within the same triangular sublattice. We use the independent set
\begin{equation}
\begin{cases}
    \boldsymbol{\delta}'_1
    &=
    \mathbf{a}_1
    =
    \frac{4a}{\sqrt{3}}\,\mathbf{e}_x,
    \\
    \boldsymbol{\delta}'_2
    &=
    \mathbf{a}_2
    =
    \frac{2a}{\sqrt{3}}\,\mathbf{e}_x
    +2a\,\mathbf{e}_y,
    \\
    \boldsymbol{\delta}'_3
    &=
    \mathbf{a}_2-\mathbf{a}_1
    =
    -\frac{2a}{\sqrt{3}}\,\mathbf{e}_x
    +2a\,\mathbf{e}_y ,
\end{cases}
\end{equation}
with the opposite vectors included implicitly in the nnn sums.

\subsection{Single-particle tight-binding Hamiltonian}

Neglecting interactions, the tight-binding Hamiltonian associated with Eq.~\eqref{eq:TBM.Hamiltonian} reads as
\begin{eqnarray}
    \hat{H}_{\mathrm{sp}}
    &=&
    -J\sum_{\langle j,\ell\rangle}
    \hat{a}_j^\dagger \hat{a}_\ell
    - \frac{\DeltaAB}{2}\sum_{j\in A}\hat{a}_j^\dagger \hat{a}_j
    + \frac{\DeltaAB}{2}\sum_{j\in B}\hat{a}_j^\dagger \hat{a}_j
    \nonumber\\
    && -J'_{\rm A}\sum_{\langle\langle j,\ell \rangle\rangle \in A}
    \hat{a}_j^\dagger \hat{a}_\ell
    -J'_{\rm B}\sum_{\langle\langle j,\ell \rangle\rangle \in B}
    \hat{a}_j^\dagger \hat{a}_\ell\,,
    \label{eq:H_sp_TB}
\end{eqnarray}
corresponding to the frist two lines of of Eq.~\eqref{eq:TBM.Hamiltonian} in the main text. After Fourier transformation, the nn and nnn contributions are expressed in terms of the form factors
\begin{equation}
\begin{cases}
    f(\mathbf{k})
    =
    \sum_{j=1}^3 e^{i\mathbf{k}\cdot\boldsymbol{\delta}_j},
    \\
    f'(\mathbf{k})
    =
    \sum_{j=1}^3
    \left(
    e^{i\mathbf{k}\cdot\boldsymbol{\delta}'_j}
    +
    e^{-i\mathbf{k}\cdot\boldsymbol{\delta}'_j}
    \right)
    =
    2\sum_{j=1}^3
    \cos\!\left(\mathbf{k}\cdot\boldsymbol{\delta}'_j\right).
\end{cases}
\end{equation}
Diagonalizing the resulting two-sublattice Hamiltonian gives
\begin{eqnarray}
    \varepsilon_{1,\pm}(\mathbf{k})
    &=&
    \pm J
    \sqrt{
    |f(\mathbf{k})|^2
    +
    \left[
    \frac{(J'_{\rm A}-J'_{\rm B})f'(\mathbf{k})+\DeltaAB}
    {2J}
    \right]^2
    }
    \nonumber\\
    &&
    -\frac{J'_{\rm A}+J'_{\rm B}}{2}
    f'(\mathbf{k})\,,
    \label{eq:TBMdispersion_app}
\end{eqnarray}
i.e., Eq.~\eqref{eq:TBMdispersion}. This is the expression used in the fits of the lowest two subbands for the nnn model.

\subsection{Limiting models: nearest neighbor and triangular}

\paragraph{Nearest neighbor model.}
The nn model is obtained by setting $J'_{\rm A}=J'_{\rm B}=0$. Eq.~\eqref{eq:TBMdispersion_app} then reduces to
\begin{equation}
    \varepsilon_{\pm}^{\textrm{nn}}(\mathbf{k})
    =
    \pm\frac{1}{2}
    \sqrt{\DeltaAB^2+4J^2|f(\mathbf{k})|^2}.
\end{equation}
In this regime, for $\DeltaAB=0$, one recovers the balanced honeycomb dispersion relation
\begin{equation}
    \varepsilon_{1,\pm}(\mathbf{k})
    =
    \pm J|f(\mathbf{k})|.
\end{equation}

\paragraph{Triangular model.}
For a large energy offset, the two sublattices effectively decouple. This is exact for $J=0$, and it is also obtained perturbatively for $\DeltaAB\gg J|f(\mathbf{k})|$. Expanding Eq.~\eqref{eq:TBMdispersion_app} in this regime gives
\begin{align}
    \varepsilon_{\textrm{A}}(\mathbf{k})
    &=
    -\frac{\DeltaAB}{2}
    -J'_{\textrm{A}}f'(\mathbf{k})
    +
    \mathcal{O}\!\left(\frac{J^2}{\DeltaAB}\right),
    \nonumber\\
    \varepsilon_{\textrm{B}}(\mathbf{k})
    &=
    +\frac{\DeltaAB}{2}
    -J'_{\textrm{B}}f'(\mathbf{k})
    +
    \mathcal{O}\!\left(\frac{J^2}{\DeltaAB}\right).
\end{align}
At leading order, these are the dispersion relations of two independent triangular lattices associated with the A and B sublattices, respectively.

\section{Obtaining the Bose--Hubbard parameters}
\label{app:Fits}

In this Appendix, we summarize the numerical procedure used to extract the parameters of the TBMs. The tunneling parameters and sublattice offset are obtained by fitting TBMs to the continuous-space single-particle band dispersion relation. The onsite interaction energies are then computed as contact-interaction matrix elements using the corresponding Wannier functions.

\subsection{Fitting the tunneling parameters}

For each value of $V_0$ and $\phig$, we diagonalize the continuous-space single-particle Hamiltonian $\hat H_0$ on a uniform grid of quasimomenta in the first Brillouin zone. The TBM dispersions introduced in Appendix~\ref{app:TBM} are then fitted to the resulting two lowest subbands.

For the nn and triangular models, the number of fitting parameters is small and the fit is performed directly by standard least-squares minimization. For the full nnn model, the four parameters $J$, $\JA$, $\JB$, $\DeltaAB$, 
can differ by several orders of magnitude, especially in the regime where the two sublattices progressively decouple. In this case, we first use a differential-evolution algorithm to generate robust initial guesses. These values are then used as starting points for a standard least-squares fit.
At the level of the differential-evolution procedure, we include a smoothness penalty to avoid spurious jumps of the fitted parameters between neighboring values of $\phig$. 

\subsection{Obtaining interaction energies}

The onsite interaction energies, $\UA$ and $\UB$ of the BHM of Eq.~\eqref{eq:TBM.Hamiltonian}, are computed from maximally localized Wannier functions associated with the lowest band manifold, which contains two subbands corresponding to the two sites of the unit cell. Starting from the Bloch eigenstates obtained by exact diagonalization of $\hat H_0$, we construct the input matrices required by \textsc{Wannier90}~\cite{Wannier90}: the Bloch eigenvalues, the overlap matrices between neighboring quasimomenta, and the projections onto localized trial orbitals centered on the A and B sites.

The Wannierization procedure yields localized orbitals $\wA(\mathbf r-\mathbf R_j)$ and $\wB(\mathbf r-\mathbf R_j)$. For contact interactions, the corresponding onsite interaction energies are then computed as
\begin{equation}
    \UAB
    =
    \tilde g
    \int d\mathbf r\,
    |\wAB(\mathbf r)|^4\,,
\end{equation}
with $\Tilde{g}$ given by Eq.~\eqref{eq:gRenorm}. The integrals are evaluated numerically on the real-space grid used for the continuous-space calculations.

\section{Density-assisted tunneling}
\label{app:DAT}

In the main text, we argued that density-assisted tunneling provides a significant correction to the nn hopping in the honeycomb lattice, even for $V_0=15\,E_r$. Here we quantify this effect using the Wannier functions obtained as described in Appendix~\ref{app:Fits}.

The density-assisted tunneling amplitude is evaluated as
\begin{equation}
    \tilde{J}\left(\tilde{g}\right)
    =
    -\tilde{g}
    \int d\mathbf{r}\,
    |\wA(\mathbf{r})|^2
    \wA^*(\mathbf{r})\wB(\mathbf{r}) .
\end{equation}
To simplify the discussion and avoid computing the density versus chemical potential over a wide range, we use below $\tilde{J}_0=\tilde{J}\left(\tilde{g}_0\right)$. Since the renormalized coupling $\tilde g$ is generally larger than $\tilde g_0$ in the parameter range considered, this gives a conservative estimate of the DAT contribution.

\begin{figure}[t]
    \centering
    \includegraphics[width=\linewidth]{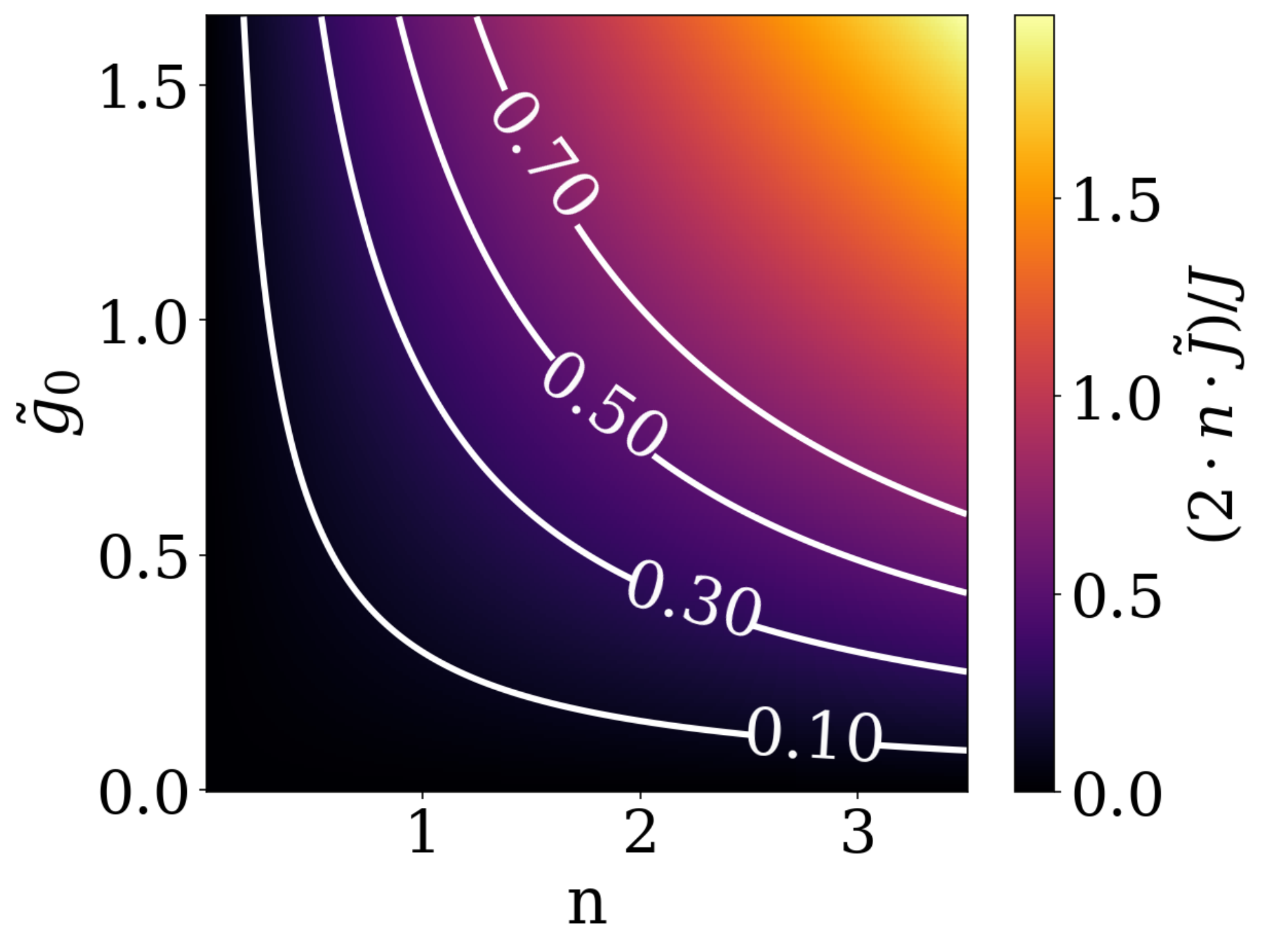}
    \caption{
    Mean-field estimate of the relative density-assisted tunneling correction,
    $2n\tilde J/J$, as a function of density $n$ and bare coupling $\tilde g_0$
    for $V_0=15\,E_r$. Contours indicate corrections of $10\%$, $30\%$, $50\%$, and $70\%$ relative to the bare nn hopping.
    }
    \label{fig:dat_app}
\end{figure}

Within the mean-field estimate used in the main text, the DAT term renormalizes the hopping according to
\begin{equation}
    J_{\rm eff}=J+2n\tilde J .
\end{equation}
The relevant dimensionless quantity for the importance of DAT corrections is the relative contribution ${2n\tilde{J}_0}/{J}$, plotted in Fig.~\ref{fig:dat_app} as a function of the density $n$ and interaction strength $\tilde g_0$, with contour lines in white.

In Fig.~\ref{fig:dat_app} we show that the relative contribution is small only at low density and weak interaction. In the region corresponding to the first Mott lobe, $MI_{1,1}$ in Fig.~\ref{fig:phasesphi0}(a), it remains modest: at the tip of the lobe it is $\sim 8.5\%$, increasing to $34.1\%$ at $\tilde{g}_0=1$. For the second lobe $MI_{2,2}$, it becomes significantly larger, reaching $\sim 30.7\%$ at the tip and $68.2\%$ at $\tilde{g}_0=1$.  
For the third lobe (as predicted by the BHM), the relative contribution is already substantial, with values of $\sim 44.0\%$ at the predicted tip and $102.3\%$ at $\tilde{g}_0=1$. For reference, the corrections at the tips of the first and second lobes predicted by the BHM are $\sim 6.8\%$ and $21.1\%$, respectively. 

This behavior explains why the standard nn BHM remains approximately valid for the first Mott lobe but fails progressively at higher filling.

%\newpage

%%%%%%%%%%%%%%%%%%%%%%%%%%%%%%%%%%%%%
% \bibliographystyle{apsrev4-2} %comment for longbibliography
\bibliography{biblioLSP,biblioDNG}

\end{document}